\documentclass[%
 aip,
 amsmath,amssymb,
 reprint,%
]{revtex4-1}

\usepackage{graphicx}
\usepackage{dcolumn}
\usepackage{bm}

\usepackage[utf8]{inputenc}
\usepackage[T1]{fontenc}
\usepackage{mathptmx}
\usepackage{etoolbox}

\makeatletter
\def\@email#1#2{%
 \endgroup
 \patchcmd{\titleblock@produce}
  {\frontmatter@RRAPformat}
  {\frontmatter@RRAPformat{\produce@RRAP{*#1\href{mailto:#2}{#2}}}\frontmatter@RRAPformat}
  {}{}
}%
\makeatother
\begin{document}

\preprint{AIP/123-QED}

\title{
Pumping by Two-frequency Steady Streaming: Impact of Amplitude Ratio and Phase}
 \author{Hyun S.\ Lee}
\affiliation{%
Department of Mathematics, University of California, Davis, Davis, CA 95616
}%
\author{Becca Thomases}%
\affiliation{ 
Department of Mathematics, Smith College, Northampton, MA 01063
}
\author{William D.\ Ristenpart}%
\affiliation{ 
Department of Chemical Engineering, University of California Davis, Davis, CA 95616
}
\author{Robert D.\ Guy}
 \email{rdguy@ucdavis.edu}
\affiliation{%
Department of Mathematics, University of California, Davis, Davis, CA 95616
}%

\date{\today}

\begin{abstract}
The classical phenomenon of steady streaming around a circular cylinder oscillating with a single frequency results in a symmetric quadrupole-like flow structure. 
In [H.S.\ Lee, \textit{et al.}, J.\ Fluid Mech.\ 1038, A34 (2026)], we demonstrated using asymptotic analysis and numerical simulations that a cylinder vibrating with two frequencies yields asymmetrical steady streaming that pumps the fluid in one direction.
The asymptotic expansion at small amplitude showed that pumping occurs at third order or higher depending on the frequency ratio, but the solutions are analytically intractable and the influence of key control parameters, like the amplitude ratio and phase difference, remained undetermined.
Here, we analyze two-frequency steady streaming on an annular domain and reduce the equations from the asymptotic expansion to a series of boundary value problems of one spatial dimension, which can be numerically solved efficiently.
For frequency ratio 2, we show that the net force on the cylinder is proportional to $\cos(\phi - \phi^*(\text{Re}))\beta/ (1+\beta^2)^{3/2}$, where $\beta$ is the amplitude ratio, $\phi$ is the phase difference, Re is the Reynolds number, and $\phi^*(\text{Re})$ is the phase that gives the largest force.
%
%
To test whether the force on an enclosed annulus serves as a useful proxy for pumping in an open system, we also performed numerical simulations of a cylinder oscillating with two-frequencies in a rectangular channel.  Our computations demonstrate that the net flux through the channel is directly proportional to the net force on the annulus over all parameter ranges tested.
Finally, we discuss the implications for using two-frequency oscillations to pump fluid in lab-on-a-chip and other devices. 
\end{abstract}

\maketitle

\section{Introduction}
Steady streaming is the nonzero mean flow generated by the nonlinear interactions of the fluid and oscillating object.\cite{riley2001steady}
While steady streaming is a classical phenomenon that has been studied for almost 150 years, \cite{strutt1884circulation} there is a resurgence of interest in steady streaming due to its use in microfluidic applications such as drug delivery, \cite{sumner2021steady} particle trapping, \cite{agarwal2018inertial, mutlu2018oscillatory, patel2014cavity, thameem2016particle, volk2020size, wang2012efficient, zhang2024particle}
bubble-driven flow, \cite{marmottant2006microfluidics, rallabandi2014two, wang2013frequency}
mixing, \cite{ahmed2009fast, huang2013acoustofluidic, kumar2011steady, liu2003acoustic} and 
pumping. \cite{huang2014reliable, marmottant2004bubble, tovar2009lateral, tovar2011lateral, zhang2024elasto} 
Almost all applications and theory involve an object oscillating with a single frequency. 
In the recent work of Lee \textit{et al.}\cite{lee2025pumping}, we show, using asymptotic analysis and numerical computation, that a circular cylinder oscillating with two frequencies generates a net force on the surrounding fluid, which produces an asymmetrical streaming in the oscillation direction and a net-directed flow.

\begin{figure*}
    \centering
    \includegraphics[width=0.9\linewidth]{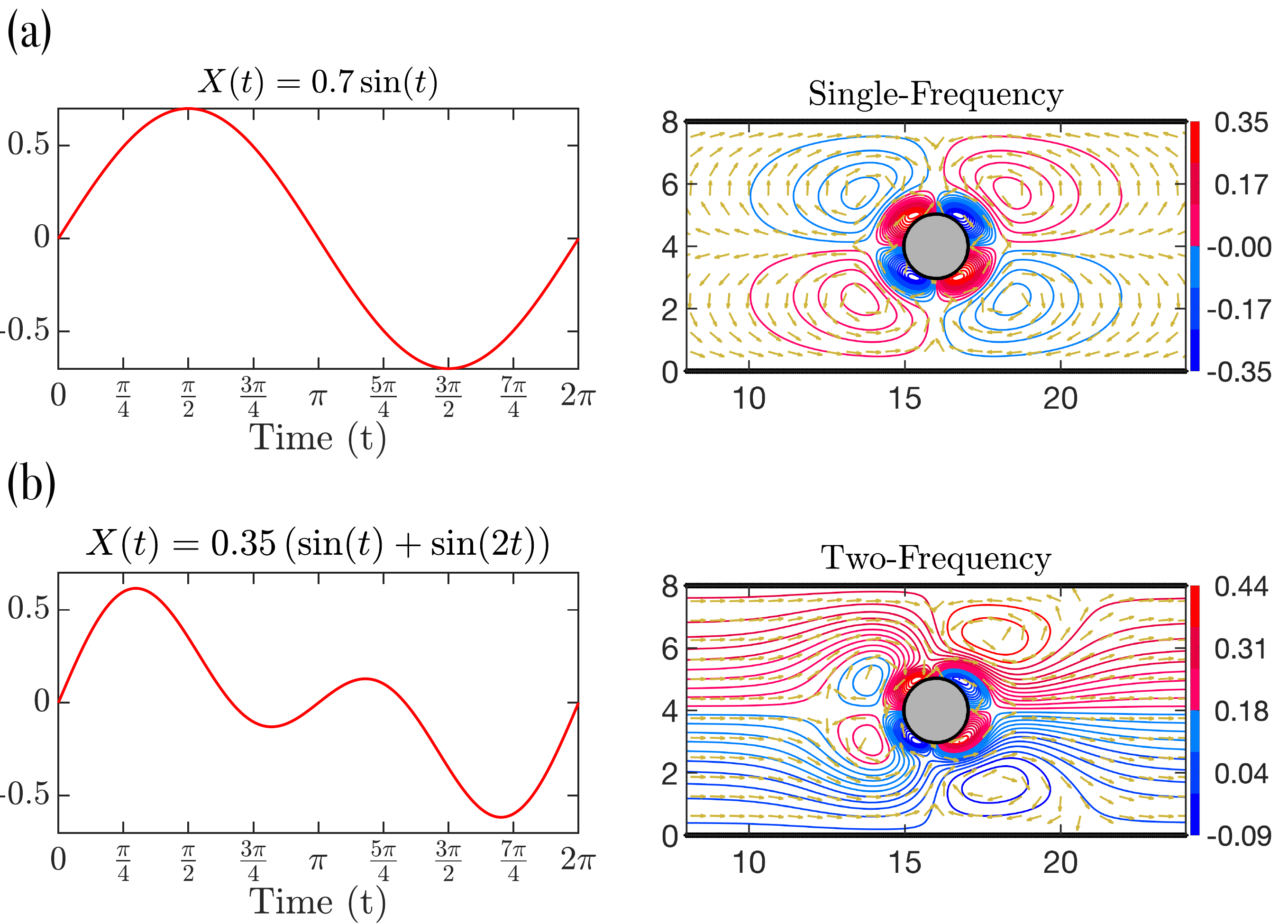}
    \caption{A cylinder is placed inside a channel of height 8 and length 32 with no-slip boundary conditions on the top and bottom and periodic conditions in the horizontal direction. At time T = 100, the streamlines and the flow directions (normalized velocity field) of the time-averaged flow are shown for the single-frequency $(\alpha = 1)$ and two-frequency $(\alpha = 2)$ oscillations for Reynolds number 50. 
The red streamlines indicate a counter-clockwise flow, and the blue is attributed to the motion in the clockwise direction. The panels on the right show the horizontal position of the cylinder's center over one period.}
\label{Fig:Introduction}
\end{figure*}

The contrast between steady streaming driven by single\-- and two-frequency oscillation is demonstrated in Figure \ref{Fig:Introduction} by the flows generated by numerical simulations. 
A cylinder is placed inside a channel with no-slip boundary conditions on the top and bottom and periodic conditions in the horizontal direction. 
The position of the cylinder's center $(X(t), 0)$, where
$X(t) = 0.35\left(\sin(t) + \sin(\alpha t)\right)$, and $\alpha = 1$ and $\alpha = 2$ corresponds to the single\-- and two-frequency case, respectively. 
For Reynolds number 50, the streamlines and the normalized velocity vectors of the time-averaged flow in Figure \ref{Fig:Introduction}(a) show that the four vortices around the cylinder maintain symmetry, and the flow away from the cylinder remain stationary.
In contrast, when the cylinder vibrates with two-frequency oscillations (Figure \ref{Fig:Introduction}(b)), the streamlines show left-right asymmetry, and there is a net horizontal flow away from the cylinder.
Details of the computational methods are provided in Appendix \ref{Sec:App5} and in Lee \textit{et al.}\cite{lee2025pumping}

The asymptotic analysis of steady streaming around a cylinder in the limit of small amplitude can be traced back to Schlichting \cite{schlichting1932berechnung} whose solution became the basis for future analyses.\cite{riley1965oscillating, stuart1966double, wang1968high, holtsmark1954boundary}
Note that past analyses have primarily focused on single-frequency oscillations, with comparatively less attention to multi-frequency oscillations.\cite{davidson1972jets, lee2025pumping}
For a cylinder vibrating with multi-frequency oscillations, Davidson \& Riley\cite{davidson1972jets} performed the analysis up to second order in amplitude and found that the resultant flow is the superposition of steady streaming flows induced by each single-frequency oscillation. 
In Lee \textit{et al.,} \cite{lee2025pumping} we showed that while steady streaming occurs at second order in amplitude, pumping is a higher-order effect. 
The solutions of the equations from the asymptotic analysis at third order and above are analytically intractable, but based on the structure of the expansion alone, we derived necessary conditions on the frequency ratio for pumping. 
Moreover, we showed that the smallest order in amplitude at which pumping occurs is 
third order for frequency ratio $\alpha = 2$.
Our work did not predict how the pumping rate varies with different parameters, which is important for microfluidic applications.


There are a variety of microchannel pumps that combine the single-frequency steady streaming and the shape asymmetry of the device to pump fluid. \cite{huang2014reliable, tovar2009lateral, tovar2011lateral}
The asymmetry illustrated in Figure \ref{Fig:Introduction}(b) is temporal rather than spatial, and pumping generated using two-frequency vibrations could lead to new designs of microfluidic pumps or potentially enhance the performance of existing designs.
Developing devices that utilize two-frequency oscillations requires understanding how the pumping rate depends on the waveform, which was not addressed in our previous work\cite{lee2025pumping} because the solutions to the asymptotic equations are analytically intractable. 
While we can quantify how the pumping rate changes in direction and strength with numerical simulations, it would be computationally expensive.

In this work, we use asymptotic analysis and numerical simulations to quantify how the streaming flow generated by two-frequency vibrations depends on amplitude ratio and phase difference of the two oscillations. 
We analyze the problem inside an annulus for frequency ratio $\alpha = 2$, and while there is no pumping inside an annular domain, a net force is produced at third order in amplitude where the solutions are analytically intractable.
Therefore, we use the solution structure from the analysis to derive series of boundary value problems in one spatial dimension and efficiently solve the equations numerically. 
Finally, we carry out numerical simulations inside a channel that contains a cylinder oscillating with two frequencies and measure the flux through the channel. 
We show that the net force from the annular domain is proportional to the flux through the channel, and thus, the size of the net force serves as a proxy for the strength of pumping.  

The sections are organized as follows. 
In Section \ref{Sec:Problem Statement}, we define the problem on an annular domain and the non-dimensionalization. 
In Section \ref{Sec:Low Amplitude Analysis}, we examine the structure of the solutions up to third order in amplitude and derive series of boundary value problems that are solved numerically. 
Moreover, we show that for frequency ratio $\alpha = 2$, the steady terms at third order produce a net force.
In Section \ref{Sec:Computational Results}, we compute the net force as a function of amplitude ratio, phase, and Reynolds number.
In Section \ref{Sec:Pumping and Net Force}, we perform computational experiments of a cylinder oscillating with two frequencies inside a channel.
We show that the net force from the analysis is proportional to the flux through the channel, and thus, the force from the analysis can be used to understand how the pumping rate varies with different parameters.

\section{Problem Statement}\label{Sec:Problem Statement}
We consider the 2D viscous flow around a cylinder that is oscillating in the horizontal direction centered at $(X(t),0)$ where 
\begin{equation}
X(t) = A_1\sin(\Omega t) + A_2\sin(\alpha\Omega t + \phi).
\label{Eq:Cylinder Motion}
\end{equation}
In Equation \eqref{Eq:Cylinder Motion}, $A_1$ and $A_2$ are the amplitudes of the two oscillations, $\Omega$ is the angular frequency of the base oscillation, $\phi$ is the phase difference, and $\alpha \geq 1$, the ratio of the two frequencies. 
We also consider the related problem of a stationary cylinder, immersed inside an oscillating horizontal far-field flow of strength, 
\begin{equation}
U(t) = -A_1\Omega\cos(\Omega t) - \alpha A_2\Omega\cos(\alpha \Omega t + \phi).
\label{Eq:Far-Field Velocity}
\end{equation}
The two formulations are related by a change in reference frame.
Switching the frame of reference from the inertial frame where the cylinder is oscillating to the non\--inertial frame in which the cylinder is fixed results in a fictitious body force, \citep{batchelor2000introduction}
\begin{equation*}
    F = -\rho U'(t) \boldsymbol{e_x},
\end{equation*}
where $\rho$ is the mass density and $\boldsymbol{e_x}$ is the direction vector in the horizontal direction. 

\begin{figure}
    \centering
    \includegraphics[width=1\linewidth]{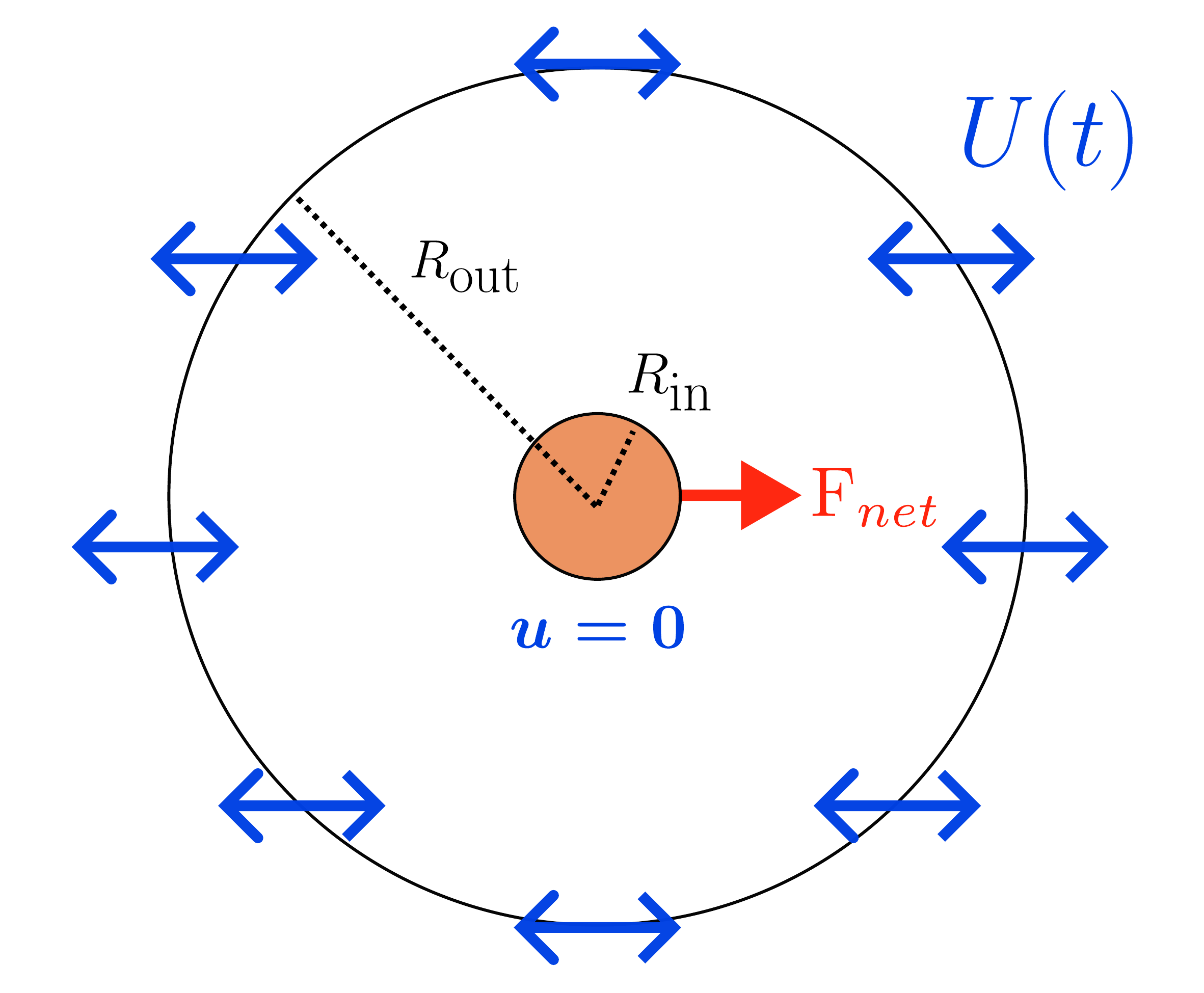}
    \caption{We solve for the flow inside an annulus where the radii of the inner and outer cylinder are $R_1$ and $R_2$, respectively. We prescribe no-slip conditions at radius $R_1$ and a horizontal velocity of strength $U(t)$ at radius $R_2$. There is no pumping inside an annulus, so we instead compute the time-averaged net force $\text{F}_{net}$ on the inner cylinder.}
    \label{Fig:Schematic}
\end{figure} 

To facilitate the analysis, we solve the problem on an annulus (Figure \ref{Fig:Schematic}) where we fix the positions of both the inner and outer cylinders of radii $R_{\text{in}}$ and $R_{\text{out}}$, correspondingly, and we impose the far-field flow of \eqref{Eq:Far-Field Velocity} around the outer cylinder.
Specifically, the boundary conditions are
\begin{equation}
u_r\left(R_{\text{in}}\right) = 0, \quad u_\theta\left(R_{\text{in}}\right) = 0,
\label{Eq:Annulus BC 1}
\end{equation}
\begin{equation}
u_r\left(R_{\text{out}}\right) = U(t)\cos(\theta), \quad u_\theta\left(R_{\text{out}}\right) = -U(t)\sin(\theta).
\label{Eq:Annulus BC 2}
\end{equation}
Steady streaming in an annulus has been analyzed previously in the case one of the inner and the outer cylinder is fixed and the other is vibrating with a single frequency.\cite{skavlem1955steady, segel1961application, bertelsen1973nonlinear, duck1979steady} The goal of our analysis is to quantify pumping from two-frequency oscillations, but there is no pumping inside an annulus, and so instead, we compute the net force on the inner cylinder.  In the later sections, we use numerical simulations to measure the flux generated by an oscillating cylinder through a channel and show that the net force computed from the asymptotic analysis is proportional to the flux through the channel.

\subsection{Non-dimensionalization}
The equations are non-dimensionalized using the characteristic length scale $R_{\text{in}}$, time scale $\Omega^{-1}$, and velocity scale $R_{\text{in}}\Omega$.
The flow around the cylinder is governed by the incompressible Navier Stokes equations:
\begin{gather}
\label{Eq:Momentum}
\text{Re} \left(\boldsymbol{u}_t + \boldsymbol{u \cdot \nabla u}\right) = -\boldsymbol{\nabla}p + \Delta \boldsymbol{u},\\
\label{Eq:Continuity}
\boldsymbol{\nabla\cdot u} = 0,
\end{gather}
where the Reynolds number is
$$\text{Re} = \frac{R_{\text{in}}^2\Omega}{\nu},$$
and $\nu$ is the kinematic viscosity.
The dimensionless amplitude ratio, $\beta$, and amplitude, $\epsilon$ are
$$\beta = \frac{A_2}{A_1},\qquad\epsilon = \frac{\sqrt{A_1^2 + A_2^2}}{R_{\text{in}}}.$$
Non-dimensionalizing the velocity $U(t)$ from Equation \eqref{Eq:Far-Field Velocity}, we obtain 
\begin{equation}
U(t) = -\frac{\epsilon}{\sqrt{1 + \beta^2}}\left(\cos(t) + \alpha \beta \cos(\alpha t + \phi)\right),
\label{Eq:NonDim Far Field}
\end{equation}
and its corresponding displacement,
\begin{equation}
X(t) = -\frac{\epsilon}{\sqrt{1 + \beta^2}}\left(\sin(t) + \beta \sin(\alpha t + \phi)\right).
\label{Eq:NonDim Displacement}
\end{equation}
In the limits of the amplitude ratio $\beta \rightarrow 0$ and $\beta \rightarrow \infty$ the displacement is 
$$\lim_{\beta \rightarrow 0}X(t) = -\epsilon\sin(t), \qquad \lim_{\beta \rightarrow \infty}X(t) = -\epsilon\sin(\alpha t + \phi),$$
and the velocity is 
$$\lim_{\beta \rightarrow 0}U(t) = -\epsilon\cos(t), \qquad \lim_{\beta \rightarrow \infty}U(t) = -\epsilon\alpha\cos(\alpha t + \phi).$$
Thus single frequency motion results in both limits, but with frequencies one and $\alpha$, correspondingly.

Finally, we define $R$ as the ratio of the two radii of the annulus,
$$R = \frac{R_{\text{out}}}{R_{\text{in}}}.$$
Thus, the boundary conditions \eqref{Eq:Annulus BC 1} and \eqref{Eq:Annulus BC 2} become
\begin{equation}
u_r(r = 1) = 0,
\label{Eq:NonDimAnnulus BC 1}
\end{equation}
\begin{equation}
u_\theta(r = 1) = 0,
\label{Eq:NonDimAnnulus BC 2}
\end{equation}
\begin{equation}
u_r(r = R) = U(t)\cos(\theta),
\label{Eq:NonDimAnnulus BC 3}
\end{equation}
\begin{equation}
u_\theta(r = R) = -U(t)\sin(\theta).
\label{Eq:NonDimAnnulus BC 4}
\end{equation}

\subsection{Non-Dimensional Regimes of the Asymptotic Analysis}

The Reynolds number, $\text{Re} = R_{\text{in}}^2\Omega / \nu$, is based on the length scale of the cylinder's radius and is related to the boundary layer thickness in which viscous forces are relevant. This so called AC boundary layer has thickness $\delta_{\text{AC}} = R_{\text{in}} /\sqrt{\text{Re}}$. \cite{raney1954acoustical, bertelsen1973nonlinear, lutz2005microscopic, chong2013inertial} Additionally, steady streaming depends on the streaming Reynolds number $(\text{Re}_s)$, which is based on the length scale of the oscillation amplitude and characterizes the Reynolds number of the streaming flow.  
The two Reynolds numbers are related by 
$$\text{Re}_s = \frac{(A_1^2 + A_2^2)\Omega}{\nu} = \epsilon^2\text{Re},$$
and the flow structure depends on both the Re and $\text{Re}_s$.
We refer to Figure 1 from Chong \textit{et al}.\cite{chong2013inertial} (which is adapted from Figure 1 of Wang \cite{wang1968high}) to illustrate the different flow regimes and associated analyses for the steady streaming problem.
For high Reynolds number, the flow exhibits a layer structure which separates the inner circulating region with the counter-rotating vortices outside the layer. 
This boundary layer is defined as the DC boundary layer, and for all Reynolds numbers, the width of the DC boundary layer $\delta_{\text{DC}}$ is greater than $\delta_{\text{AC}}$.\cite{raney1954acoustical, bertelsen1973nonlinear, lutz2005microscopic, chong2013inertial}
Concurrently, if the streaming Reynolds number is large, the outer flow also exhibits a boundary layer, which was analyzed by Riley \cite{riley1965oscillating} and Stuart. \cite{stuart1966double}
At low Reynolds number, there are no boundary layers, and Holtsmark \textit{et al.} \cite{holtsmark1954boundary} analyzed the problem as a regular perturbation.
Though as noted in Chong \textit{et al}.\cite{chong2013inertial}, the regular perturbation solution from Holtsmark \textit{et al.}\cite{holtsmark1954boundary} contains the single boundary layer solutions from Schlichting \cite{schlichting1932berechnung} and Wang \cite{wang1968high}, and is thus appropriate provided that the streaming Reynolds number is small.
 In this paper, we focus on the low streaming Reynolds number regime $(\text{Re}_s \ll 1)$ so that the problem can be analyzed as a regular perturbation.

\section{Low Amplitude Analysis}\label{Sec:Low Amplitude Analysis}
We examine the problem in the limit of small amplitude ($\epsilon \ll 1$) and restrict the analysis to small streaming Reynolds number ($\text{Re}_s \ll 1$) to solve the problem as a regular perturbation. 
In Lee \textit{et al}.\cite{lee2025pumping}, we used the structure of the regular perturbation analysis and derived necessary conditions on the frequency ratio $\alpha$ to produce a net force.   
Moreover, we showed that the lowest order in amplitude at which pumping occurs is third order for frequency ratio $\alpha = 2$, and thus, the largest pumping is generated for frequency ratio $\alpha = 2$.
In this paper, we analyze the problem for general amplitude ratio $\beta$ and phase $\phi$ at $\alpha = 2$ for varying Reynolds numbers.
Thus, we carry out the asymptotic analysis to third order and compute the terms that generate a net force.  

\subsection{Governing Equations}
We analyze the problem in terms of the stream function, $\psi$, which is related to the velocity by
\begin{equation}
u_r = \frac{1}{r}\frac{\partial \psi}{\partial \theta},\quad u_\theta = -\frac{\partial \psi}{\partial \theta},
\label{Eq: Stream Function Def}
\end{equation}
and
$$\boldsymbol{\nabla}\times\boldsymbol{u} = -\Delta \psi.$$
We take the curl of the momentum equation \eqref{Eq:Momentum} to obtain
\begin{equation}
\left(\Delta -\text{Re}\frac{\partial }{\partial t}\right) \Delta \psi = \text{Re}\; \boldsymbol{u \cdot \nabla}\left(\Delta \psi\right).
\label{Eq:Momentum Stream}
\end{equation}
The boundary conditions of \eqref{Eq:NonDimAnnulus BC 1}-\eqref{Eq:NonDimAnnulus BC 4} in terms of the stream function $(\psi)$ are 
\begin{equation}
\label{Eq:Inner BC Stream Function 1}
\psi(r = 1) = 0,
\end{equation}
\begin{equation}
\label{Eq:Inner BC Stream Function 2}
\frac{\partial \psi}{\partial r}(r = 1) = 0,
\end{equation}
\begin{equation}
\label{Eq:Far-Field Part 1}
\left.\psi\right|_{r = R}= -\frac{\epsilon R}{\sqrt{1+\beta^2}}\left(\cos(t) + 2\beta\cos(2t + \phi)\right)\sin(\theta),
\end{equation}
\begin{equation}
\label{Eq:Far-Field Part 2}
\left.\frac{\partial \psi}{\partial r}\right|_{r = R} = -\frac{\epsilon}{\sqrt{1+\beta^2}}\left(\cos(t) + 2\beta\cos(2t + \phi)\right)\sin(\theta).
\end{equation}
We consider the low amplitude limit and expand the solution in terms of $\epsilon = \sqrt{A_1^2 + A_2^2}/R_{\text{in}}$:
 \begin{equation}
\label{Eq:Stream Function Expansion}
\psi = \epsilon\psi_1 + \epsilon^2\psi_2 + \epsilon^3\psi_3 + O\left(\epsilon^4\right).
\end{equation}
Substituting the expansion \eqref{Eq:Stream Function Expansion} into \eqref{Eq:Momentum Stream}-\eqref{Eq:Far-Field Part 2}, we obtain successive equations at each order:
\begin{gather}
\textbf{First Order}\nonumber\\
\label{Eq:First Order Line 1}
\left(\Delta - \text{Re}\frac{\partial}{\partial t}\right) \Delta \psi_1 = 0,\\
\label{Eq:First Order Line 2}
\psi_1(r = 1) = 0, \quad \frac{\partial \psi_1}{\partial r}(r = 1) = 0,\\
\psi_1(r = R) = -\frac{R}{\sqrt{1+\beta^2}}\left(\cos(t)+ 2\beta\cos(2t+\phi)\right)\sin(\theta),\label{Eq:First Order Line 3 i}\\
\frac{\partial \psi_1}{\partial r}(r = R) = -\frac{1}{\sqrt{1+\beta^2}}\left(\cos(t) + 2\beta\cos(2t+\phi)\right)\sin(\theta),\label{Eq:First Order Line 3 ii}
\end{gather}
\begin{gather}
\textbf{Second Order}\nonumber\\
\label{Eq:Second Order Line 1}
\left(\Delta - \text{Re}\frac{\partial}{\partial t}\right) \Delta \psi_2 = \text{Re}\;\boldsymbol{u}_1\boldsymbol{\cdot \nabla}\left(\Delta \psi_1\right),\\
\label{Eq:Second Order Line 2}
\psi_2(r = 1) = 0, \quad \frac{\partial \psi_2}{\partial r}(r = 1) = 0,\\
\label{Eq:Second Order Line 3}
\psi_2(r = R) = 0, \quad \frac{\partial \psi_2}{\partial r}(r = R) = 0,
\end{gather}
\begin{gather}
\textbf{Third Order}\nonumber\\
\left(\Delta - \text{Re}\frac{\partial}{\partial t}\right) \Delta \psi_3 = \text{Re}\;\boldsymbol{u}_2\boldsymbol{\cdot \nabla}\left(\Delta \psi_1\right),\nonumber\\
+ \text{Re}\;\boldsymbol{u}_1\boldsymbol{\cdot \nabla}\left(\Delta \psi_2\right),\label{Eq:Third Order Line 1}\\
\label{Eq:Third Order Line 2}
\psi_3(r = 1) = 0, \quad \frac{\partial \psi_3}{\partial r}(r = 1) = 0,\\
\label{Eq:Third Order Line 3}
\psi_3(r = R) = 0, \quad \frac{\partial \psi_3}{\partial r}(r = R) = 0.
\end{gather}

\subsection{Solution Structure}
At first order in amplitude, the boundary conditions \eqref{Eq:First Order Line 3 i} and \eqref{Eq:First Order Line 3 ii} define the solution structure as 
\begin{equation}
\label{Eq:First Order Solution}
\psi_1 = \Re\left(a_1(r)e^{-it} + a_2(r)e^{-i2t}\right)\sin(\theta),
\end{equation}
where $\Re(z)$ denotes the real part of $z$.
Substituting the first order solution \eqref{Eq:First Order Solution} into the right-hand side of Equation \eqref{Eq:Second Order Line 1}, which is a quadratic nonlinearity, the products of $e^{\pm it}$ and $e^{\pm i2t}$ results in a steady term and terms proportional to $\{e^{\pm it }, e^{\pm i2t }, e^{\pm i3t }, e^{\pm i4t }\}$. 
Therefore, at second order, we obtain 
\begin{equation}
\label{Eq:Second Order Solution}
\psi_2 = \Re\left(\sum_{k = 1}^{4}b_k(r)e^{-ikt}\right)\sin(2\theta) + b_0(r)\sin(2\theta).
\end{equation}
Similarly, we substitute \eqref{Eq:First Order Solution} and \eqref{Eq:Second Order Solution} into the quadratic nonlinearity of \eqref{Eq:Third Order Line 1}, and the products of terms proportional to $\{e^{\pm it}, e^{\pm i2t}\}$ and $\{1, e^{\pm it }, e^{\pm i2t }, e^{\pm i3t }, e^{\pm i4t }\}$ result in a solution structure of \eqref{Eq:Third Order Line 1}-\eqref{Eq:Third Order Line 3} as
\begin{eqnarray}
\psi_3 &=& \Re\left(\sum_{k = 1}^{6}c_k(r)e^{-ikt}\right)\sin(3\theta)+ c_0(r)\sin(3\theta)\nonumber\\
&&+ \Re\left(\sum_{k = 1}^{6}d_k(r)e^{-ikt}\right)\sin(\theta)+ d_0(r)\sin(\theta).\label{Eq:Third Order Solution}
\end{eqnarray}
Note that in  \eqref{Eq:First Order Solution}-\eqref{Eq:Third Order Solution}, the dependence on time and angular coordinate, $\theta$, is known, and only the radial dependence is unknown. We derive a series of one dimensional boundary value problems to determine how the solution depends on the radial variable, $r$.

We begin by defining the differential operator, 
\begin{equation}
D_\ell = \frac{d^2}{dr^2} + \frac{1}{r}\frac{d}{dr} - \frac{\ell^2}{r^2}.
\label{Eq: Differential Operator}
\end{equation}
Substituting the first order solution \eqref{Eq:First Order Solution} into Equations \eqref{Eq:First Order Line 1}-\eqref{Eq:First Order Line 3 ii}, we obtain
\begin{gather}
  \textbf{First Order} \nonumber  \\
  D_1\left(D_1 + ik\;\text{Re}\right)a_k = 0, \label{Eq:Ak}\\
  a_k(1) = 0, \quad a_k'(1) = 0,\label{Eq:Ak Bd 1}
\end{gather}
for $k = 1,2$.
The boundary conditions at $r = R$ are
\begin{equation}
\label{Eq:Ak Bd 2}
a_1(R) = -\frac{R}{\sqrt{1 + \beta^2}}, \quad a_1'(R) = -\frac{1}{\sqrt{1 + \beta^2}},
\end{equation}
and 
\begin{equation}
\label{Eq:Ak Bd 3}
a_2(R) = -\frac{R\; 2\beta e^{-i\phi}}{\sqrt{1 + \beta^2}}, \quad a_2'(R) = -\frac{2\beta e^{-i\phi}}{\sqrt{1 + \beta^2}}.
\end{equation}
For the equations at second order, we substitute the second order solution \eqref{Eq:Second Order Solution} into Equations \eqref{Eq:Second Order Line 1}-\eqref{Eq:Second Order Line 3} and obtain
\begin{gather}
  \textbf{Second Order} \nonumber  \\
  D_2\left(D_2 + ik\;\text{Re}\right)b_k(r) = \Omega_k(r),\label{Eq:Bk}\\
  b_k(1) = 0, \quad b_k'(1) = 0, \label{Eq:Bk Bd 1}\\
  b_k(R) = 0, \quad b_k'(R) = 0. \label{Eq:Bk Bd 2}
\end{gather}
Similarly, we substitute the third order solution \eqref{Eq:Third Order Solution} into Equations \eqref{Eq:Third Order Line 1}-\eqref{Eq:Third Order Line 3} to derive the equations at third order as
\begin{gather}
  \textbf{Third Order} \nonumber  \\
  D_3\left(D_3 + ik\;\text{Re}\right)c_k(r) = \rho_k(r),\label{Eq:Ck}\\
  c_k(1) = 0, \quad c_k'(1) = 0, \label{Eq:Ck Bd 1}\\
  c_k(R) = 0, \quad c_k'(R) = 0, \label{Eq:Ck Bd 2}
\end{gather}
\begin{gather}
  D_1\left(D_1 + ik\;\text{Re}\right)d_k(r) = \xi_k(r),\label{Eq:Dk}\\
  d_k(1) = 0, \quad d_k'(1) = 0, \label{Eq:Dk Bd 1}\\
  d_k(R) = 0, \quad d_k'(R) = 0. \label{Eq:Dk Bd 2}
\end{gather}
The terms $\Omega_k(r)$, $\rho_k(r)$, and $\xi_k(r)$ are defined in Appendix \ref{Sec:App1}.
 While the equations at first and second order can be solved analytically \cite{holtsmark1954boundary, chong2013inertial}, equations \eqref{Eq:Ck}-\eqref{Eq:Dk Bd 2} at third order are analytically intractable. Therefore, we solve the boundary value problems numerically, which we describe in detail in Appendix \ref{Sec:App3}.

In Figure \ref{Fig:Flow Structure}, we show the contours of the stream function for single- and two-frequency ($\beta = 1$) oscillations for phase difference $\phi = 0$, $\epsilon = \sqrt{0.02}\approx0.14$, $R = 8$, and Re = 50. The streamlines for the single-frequency case ($\beta = 0$) show the classic quadrupole-like flow structure and exhibit symmetry across the lines $y = \pm x$. For the two-frequency case, like the flow through the channel shown in Figure \ref{Fig:Introduction}(b), the streamlines in the annular domain are no longer left-right symmetric. Unlike the channel, there is no pumping through this closed domain, and we show in the following section that a net force is produced on the cylinder. 
\begin{figure}
    \centering
    \includegraphics[width=1\linewidth]{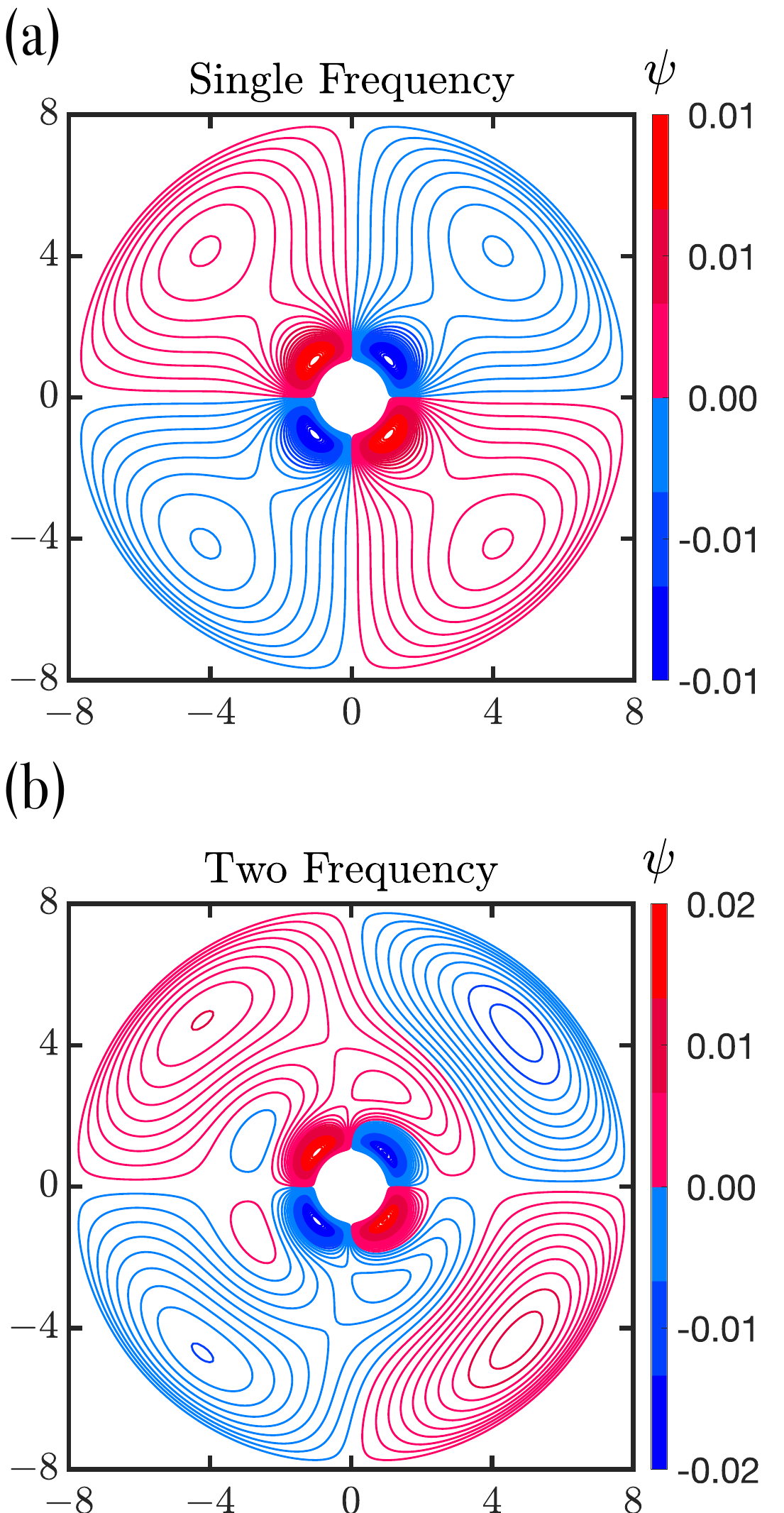}
    \caption{Contours of the steady stream function for (a) single-frequency and (b) two-frequency ($\beta = 1$) oscillations for phase difference $\phi = 0$, $\epsilon = \sqrt{0.02}\approx0.14$, $R = 8$, and Re = 50.
}
    \label{Fig:Flow Structure}
\end{figure} 

\subsection{Net Force}
The interaction of the inner cylinder and the flow generates a net force, and we show the steady term proportional to $\sin(\theta)$ that appears in the third-order solution \eqref{Eq:Third Order Solution} produces a net force. 
Assume a steady solution of the form,
$$\psi^s = \sum_{k = 1}^{\infty}f_k(r)\sin(k\theta).$$
The pressure can be computed using the momentum equation \eqref{Eq:Momentum} to obtain
$$p^s = p_\infty + \sum_{k = 1}^\infty g_k(r)\cos(k\theta).$$
The traction force in the horizontal direction is
\begin{equation}
    \left(\sigma\cdot \hat{e}_r\right)\cdot \hat{e}_x = \sigma_{rr}\cos(\theta) - \sigma_{r\theta}\sin(\theta),   
\label{Eq:Surface Stress}
\end{equation}
where $\hat{e}_{r}$ and $\hat{e}_x$ are the direction vectors for the $r$- and $x$-directions, respectively, and $\sigma$ is the stress tensor.
Using the boundary conditions on the inner cylinder from Equation \eqref{Eq:NonDimAnnulus BC 1}, the components of the stress on the cylinder are
\begin{equation*}
    \sigma_{rr} = -p_{\infty} - \sum_{k=1}^{\infty}g_k(1)\cos(k\theta), \quad \sigma_{r\theta} = -\sum_{k=1}^{\infty}f_k''(1)\sin(k\theta).
\end{equation*}
Integrating the surface stress \eqref{Eq:Surface Stress} over the inner cylinder, all the terms corresponding to $k \geq 2$ vanish due to symmetry, and the net horizontal force is 
\begin{gather}
\text{F}_{net} = \int_0^{2\pi} \sigma_{rr}\cos(\theta) - \sigma_{r\theta}\sin(\theta)\;d\theta \nonumber\\
 = \pi\left(-g_1(1) + f_1''(1)\right).    \label{Eq:Net Force}
\end{gather}
Therefore, the term proportional to $\sin(\theta)$ that appears at third order in the solution \eqref{Eq:Third Order Solution} results in a net force.

We solve Equations \eqref{Eq:Dk}-\eqref{Eq:Dk Bd 2} with $k = 0$ and its corresponding pressure to compute the net force.  As shown in Appendix \ref{Sec:App1}, the term $\xi_0$  of Equation \eqref{Eq:Dk} results from products involving $\{a_1(r), a_2(r), b_1(r), b_2(r)\}$ through the quadratic nonlinearity. Thus computing $d_0(r)$, requires the solutions of Equations \eqref{Eq:Ak}-\eqref{Eq:Bk Bd 2} for $k=1,2$. In Appendix \ref{Sec:App2}, we show that the steady pressure at third order can be computed similarly by substituting our velocity components into the momentum equation \eqref{Eq:Momentum}.

\section{Net Force as a Function of Amplitude Ratio, Phase, and Reynolds Number}
\label{Sec:Computational Results}
In the previous section, we showed for frequency ratio $\alpha = 2$, that the steady term proportional to $\sin(\theta)$ at third order in amplitude produces a net force. 
Therefore, the net force scales like $\epsilon^3$, i.e., 
$\text{F}_{net} = \epsilon^3\tilde{\text{F}}_{net},$ where $\tilde{\text{F}}_{net}$ is $O(1)$. 
Note that the cubic scaling in amplitude for the net force is independent of other parameters. When we display the net force for amplitude ratio, phase, and Reynolds number in the following sections, we neglect the $\epsilon^3$ in the force, and thus, we plot 
$$\tilde{\text{F}}_{net} = \epsilon^{-3}\text{F}_{net},$$
as the net force.
In the numerical computation of the net force, we use 120 Chebyshev points on a domain size of $R = 8$. In Appendix \ref{Sec:App6}, we consider other domain sizes.

\subsection{Amplitude Ratio ($\beta$) and Phase Difference ($\phi$)} \label{Sec:Amplitude and Phase} 
\begin{figure}
    \centering
    \includegraphics[width=1\linewidth]{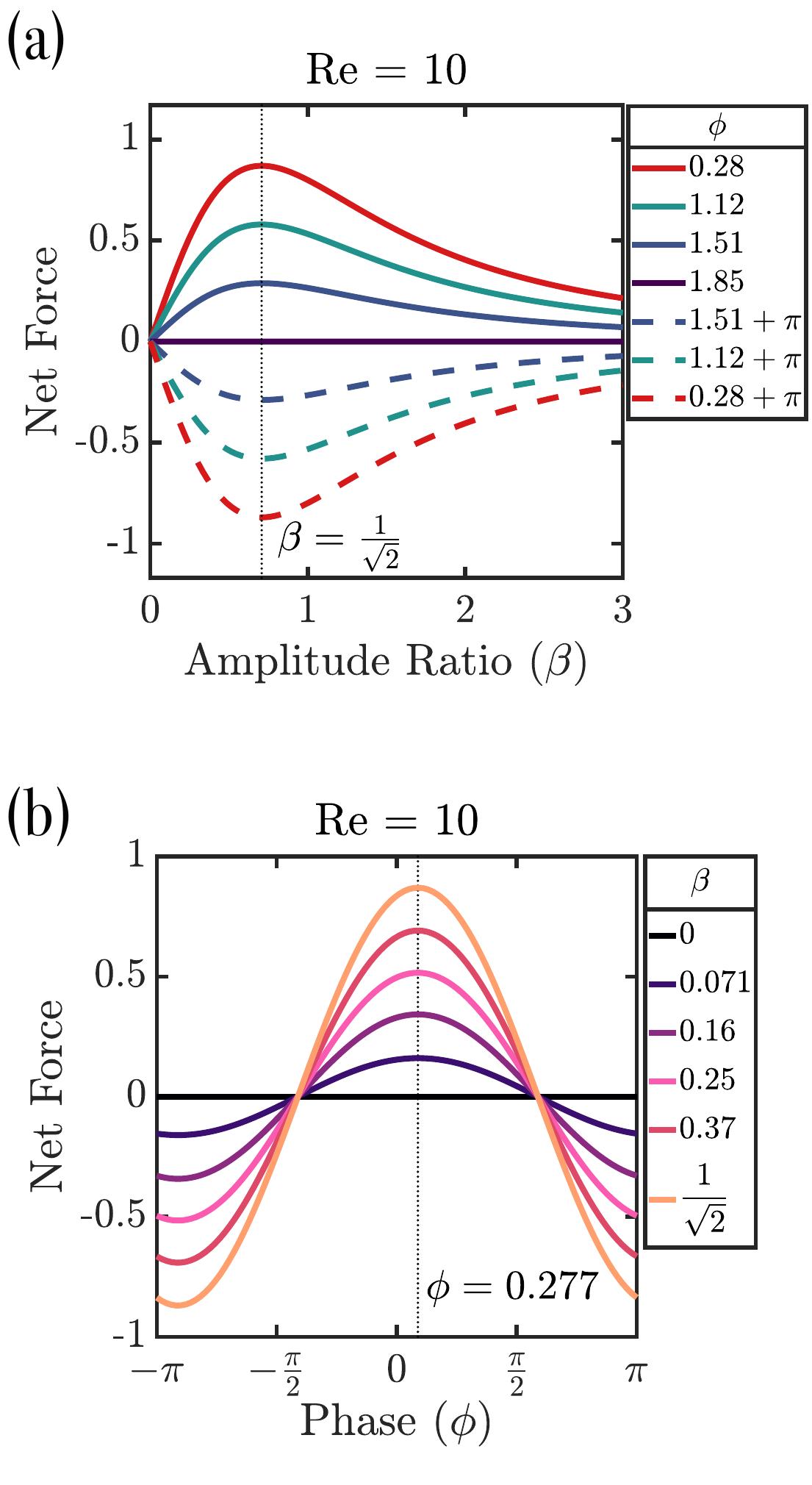}
    \caption{For Re = 10, net force as a function of (a) amplitude ratio $0\leq \beta \leq 3$ for varying phase and (b) phase $-\pi\leq \phi \leq \pi$ for varying amplitude ratio. The vertical dashed lines show $\beta_{max} = 1/\sqrt{2}$ and $\phi_{max}=0.277$ for (a) and (b) respectively.}
\label{Fig:Net Force Beta and Phi}
\end{figure} 

Figure \ref{Fig:Net Force Beta and Phi}(a) shows the net force at Re = 10 for amplitude ratio $0 \leq \beta \leq 3$ for different phases. As expected, the net force is zero at $\beta = 0$ and appears to approach zero for large $\beta$ because there is no net force for single-frequency oscillations. Because of these limits, there must be a value of the amplitude ratio that corresponds to the peak force for each phase difference. What is notable is that the location of the extremum occurs at the same amplitude ratio, independent of the phase. In the next section we show that the extremum is achieved at $\beta = 1/\sqrt{2}$.

The data also show that the force can be positive or negative, depending on the phase. We note that the value of the force at phases $\phi$ and $\phi+\pi$ are the negative of one another. We highlight this symmetry between the net force and its corresponding force at a phase shifted by $\Delta\phi=\pi$ with solid and dash dotted-lines in the figure. In the next section, we show how the symmetry results from the form of the asymptotic equation. 

Figure \ref{Fig:Net Force Beta and Phi}(b) shows the net force as a function of phase $-\pi \leq \phi \leq \pi$ for different amplitude ratios. The curves appear to have a sinusoidal form with a maximal net force at approximately $\phi \approx 0.277$ independent of the amplitude ratio. 

\begin{figure}
    \centering
    \includegraphics[width=0.9\linewidth]{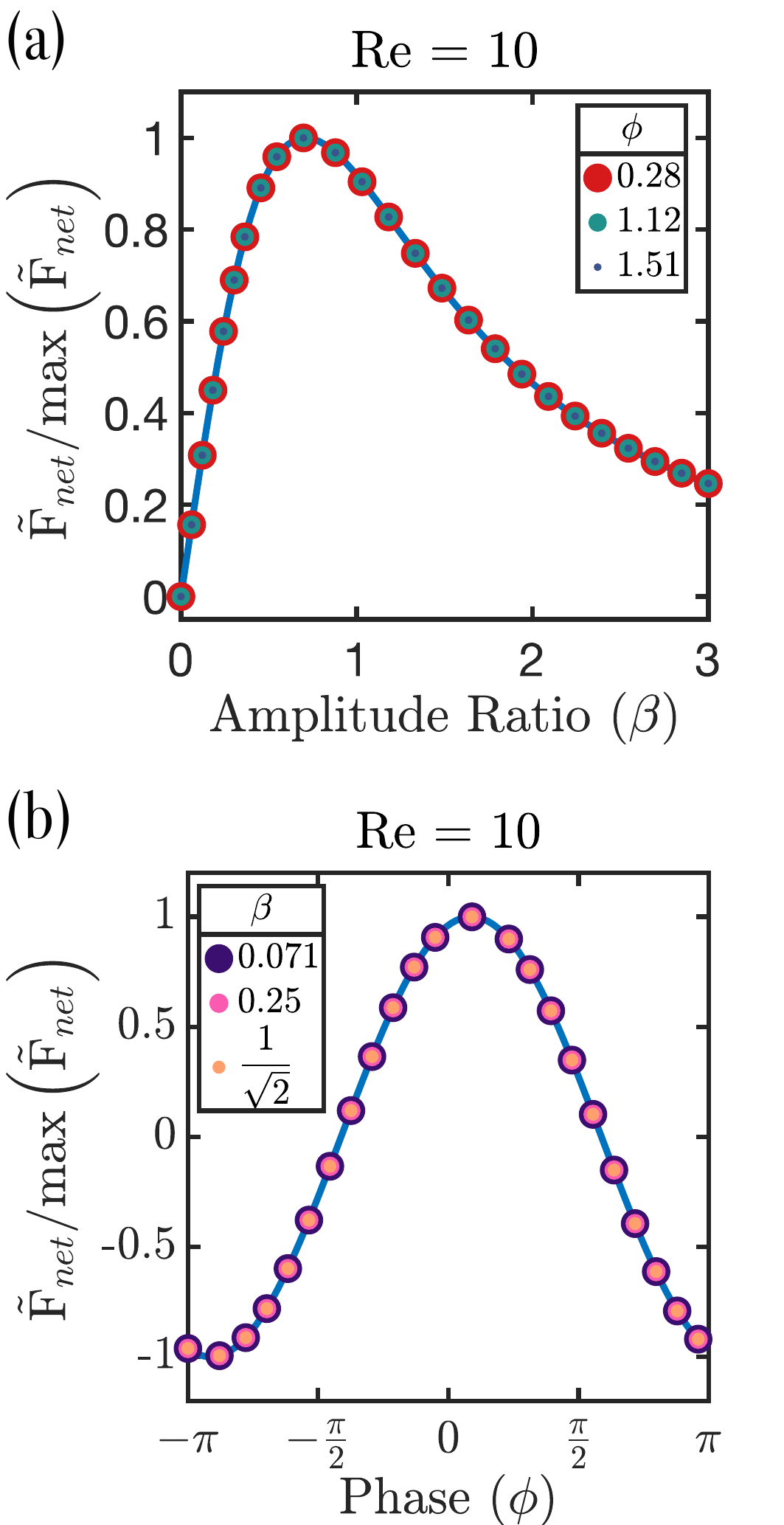}
    \caption{Scaled by the maximum value, the net force collapses onto a single curve as a function of (a) amplitude ratio $0\leq\beta\leq 3$ and (b) phase $-\pi\leq\phi\leq \pi$. The solid blue lines represent the analytical functions (a) $f(\beta)$ from \eqref{Eq:Value of Function f} and (b) $h(\phi)$ from \eqref{Eq:Value of Function h}.}
\label{Fig:Net Force Beta and Phi Similar}
\end{figure}

The different curves in Figure \ref{Fig:Net Force Beta and Phi}(a,b) appear to have the same basic shape. In Figure \ref{Fig:Net Force Beta and Phi Similar}(a,b) we normalize the corresponding data by their maximum values, and show that the normalized data collapse onto a single curve. In the next subsection, we derive a form for the dependence on the amplitude ratio, $\beta$, by revisiting the asymptotic analysis.    
We confirm the sinusoidal dependence on phase by fitting the normalized data with a single Fourier mode (fitting error is below $10^{-8}$). The resulting fit is 
\begin{equation}
h(\phi) = \cos(\phi - \phi^*),
\label{Eq:Value of Function h}
\end{equation} 
where $\phi^* \approx 0.277$ for Re = 10, and we later analyze the dependence of $\phi^*$ on Re.

\subsubsection{Force Scaling with Amplitude Ratio}\label{Sec:Amplitude Scaling}
We aim to understand how the net force depends on the amplitude ratio $\beta$ by revisiting the low amplitude analysis of Section \ref{Sec:Low Amplitude Analysis}.
At third order in amplitude, the term involving $d_0(r)$ in the expression of $\psi_3$ of \eqref{Eq:Third Order Solution} is solely responsible for the net force. The scale of $d_0(r)$ with amplitude depends on the scale of the right-hand side of \eqref{Eq:Dk}, which involves a product of lower order terms. Thus, we begin by examining how the first and second order solutions, $\psi_1$ and $\psi_2$, change with amplitude ratio $\beta$.   

At first order, the scale of $a_1(r)$ and $a_2(r)$ in the expression of $\psi_1$ from \eqref{Eq:First Order Solution} is determined by the boundary condition \eqref{Eq:Ak Bd 2} and \eqref{Eq:Ak Bd 3}. To incoorporate the scaling of these terms with amplitude ratio, we express the first order soltuion as 
$$\psi_1 = \Re\left(C_1\tilde{a}_1(r)e^{-it} + C_2\tilde{a}_2(r) e^{-i2t}\right)\sin(\theta),$$
where 
\begin{equation}
C_1 = \frac{1}{\sqrt{1 + \beta^2}}, \qquad C_2 = \frac{\beta}{\sqrt{1 + \beta^2}},
\label{Eq:scale_factors}
\end{equation}
account for the scaling.

At second order, we compute how each term $b_k(r)$ scales in \eqref{Eq:Second Order Solution} by noting that the right-hand side of \eqref{Eq:Bk}, $\Omega_k(r)$, is obtained from the products of derivatives of $\psi_1$. We rewrite the second order solution \eqref{Eq:Second Order Solution} as
\begin{eqnarray*}
\psi_2 &=& \Re\left(\sum_{k = 1}^{4}b_k(r)e^{-ikt}+b_0^{(1)}(r) + b_0^{(2)}(r)\right)\sin(2\theta)\\
&=&\Re\left(C_1^2\left(\tilde{b}_0^{(1)}(r) + \tilde{b}_2(r)e^{-i2t}\right) + C_2^2\left(\tilde{b}_0^{(2)}(r) + \tilde{b}_2(r)e^{-i4t}\right)\right.\\
&&\left.+ C_1C_2\left(\tilde{b}_1(r)e^{-it}+\tilde{b}_3(r)e^{-i3t}\right)\right)\sin(2\theta).
\end{eqnarray*}
This expression for $\psi_2$ with three pairs of terms grouped together by amplitude ratio dependence highlights how the different scalings arise. The first two pairs involving a steady term and an unsteady term correspond to the single-frequency solution from each of the two driving frequencies, and thus, the pairs are proportional to $C_1^2$ and $C_2^2$, correspondingly. The last two unsteady terms arise from the interaction between the driving frequencies, resulting in the scale factor $C_1$$C_2$.

At third order the scaling of each term arises from products of the terms in $\psi_1$ and $\psi_2$ through the quadratic nonlinearity.  Continuing the computation to third order we express $\psi_3$ as 
$$\psi_3 = C_1^2C_2\left(\tilde{d}_0^{\;(1)}(r) e^{-it}e^{it} + \tilde{d}_0^{\;(2)}(r) e^{-i2t}e^{i2t}\right)\sin(\theta) + \cdots,$$
where, for simplicity, we omit the terms that have no contribution to the net force.
Therefore $d_0(r)$ in the expression of \eqref{Eq:Third Order Solution} is proportional to $C_1^2C_2$, and thus, the net force from viscous stress scales like $C_1^2C_2$. As shown in Appendix \ref{Sec:App2}, the pressure contribution to the net force has the same scaling.

From this scaling, we define the normalized amplitude ratio dependence of the force  
as
\begin{equation}
f(\beta) = \frac{C_1^2C_2}{\max_{\beta}\left(C_1^2C_2\right)} =\frac{3\sqrt{3}\beta}{2\left(1+\beta^2\right)^{3/2}},
\label{Eq:Value of Function f}
\end{equation}
which is plotted in in 
Figure \ref{Fig:Net Force Beta and Phi Similar}(a) along with the normalized data from the numerical computation of the force. This function achieves its maximum value of 1 at $\beta = 1/\sqrt{2}$, and limits to zero as $\beta\rightarrow 0$ and $\beta\rightarrow\infty$.


\subsubsection{Reversal of the Direction of the Force} \label{Sec:Computational Results Reversal}
We analytically show that the resulting forces generated by oscillation phase difference $\phi$ and $\phi+\pi$ are in the opposite directions and of the same strength, as suggested by the computational results in Figure \ref{Fig:Net Force Beta and Phi}(a). We denote the horizontal velocity on the outer cylinder for amplitude ratio $\beta$ and phase difference $\phi$ by $U(t;\beta,\phi)$.   We return to the velocity on the boundary from equation \eqref{Eq:NonDim Far Field} to examine the effect of a change in phase difference by $\pi$:
\begin{align*}
U(t;\beta,\phi+\pi) & = -\frac{\epsilon}{\sqrt{1 + \beta^2}}\left(\cos(t) + \alpha \beta \cos(\alpha t + \phi +\pi)\right) \\
              &= -\frac{\epsilon}{\sqrt{1 + \beta^2}}\left(\cos(t) - \alpha \beta \cos(\alpha t + \phi)\right) \\
              & = U(t;-\beta,\phi).
\end{align*}
Thus a change of the phase difference by $\pi$ is equivalent to replacing $\beta$ with $-\beta$. From the previous section, the net force scale like $C_{1}^2C_{2}$ where the scale factors are defined in equation \eqref{Eq:scale_factors}. A change in sign in $\beta$ results in a change in sign of $C_{2}$. Therefore a change in phase difference of $\pi$ results in a force of the same strength but of opposite sign. 


\subsection{Net Force for Varying Reynolds Numbers}
\begin{figure}
    \centering
    \includegraphics[width=0.9\linewidth]{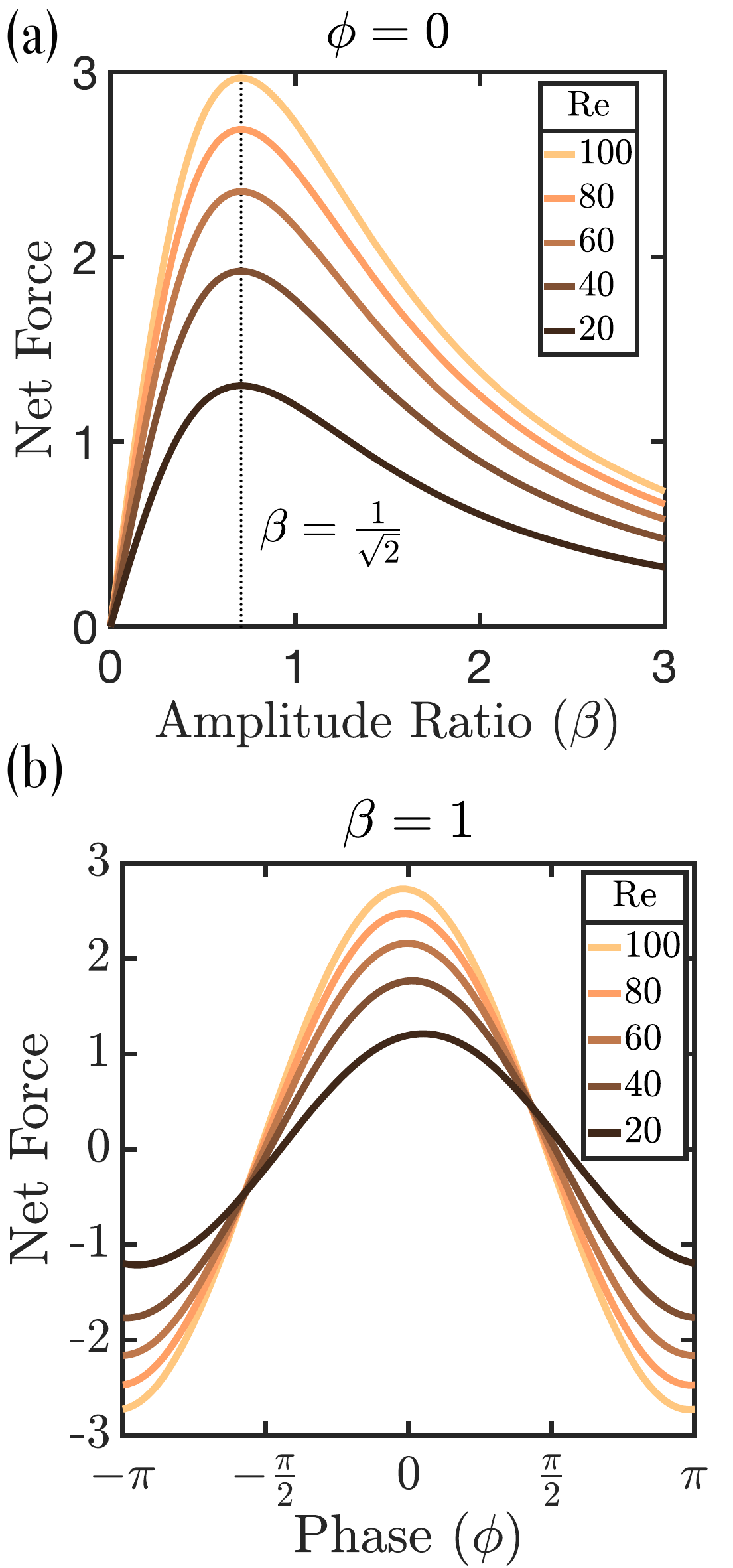}
    \caption{For varying Reynolds numbers, net force as a function of (a) amplitude ratio $0 \leq \beta \leq 3$ at $\phi = 0$ and (b) phase $-\pi \leq \phi \leq \pi$ at $\beta = 1$. The vertical dashed line in (a) shows $\beta_{max} = 1/\sqrt{2}$.} 
\label{Fig:Net Force Reynolds Number}
\end{figure}  
Figure \ref{Fig:Net Force Reynolds Number} shows the net force as a function of amplitude ratio $\beta$ and phase $\phi$ for various Reynolds numbers. In Figure \ref{Fig:Net Force Reynolds Number}(a), we compute the force for amplitude ratio $0 \leq \beta \leq 3$ at a fixed phase $\phi = 0$, which shows that the net force increases with Reynolds number. As expected from the the analysis of the previous section, the net force achieves its maximum at $\beta = 1/\sqrt{2}$, and is proportional to $f(\beta)$ from \eqref{Eq:Value of Function f}.

\begin{figure}
    \centering
    \includegraphics[width=0.9\linewidth]{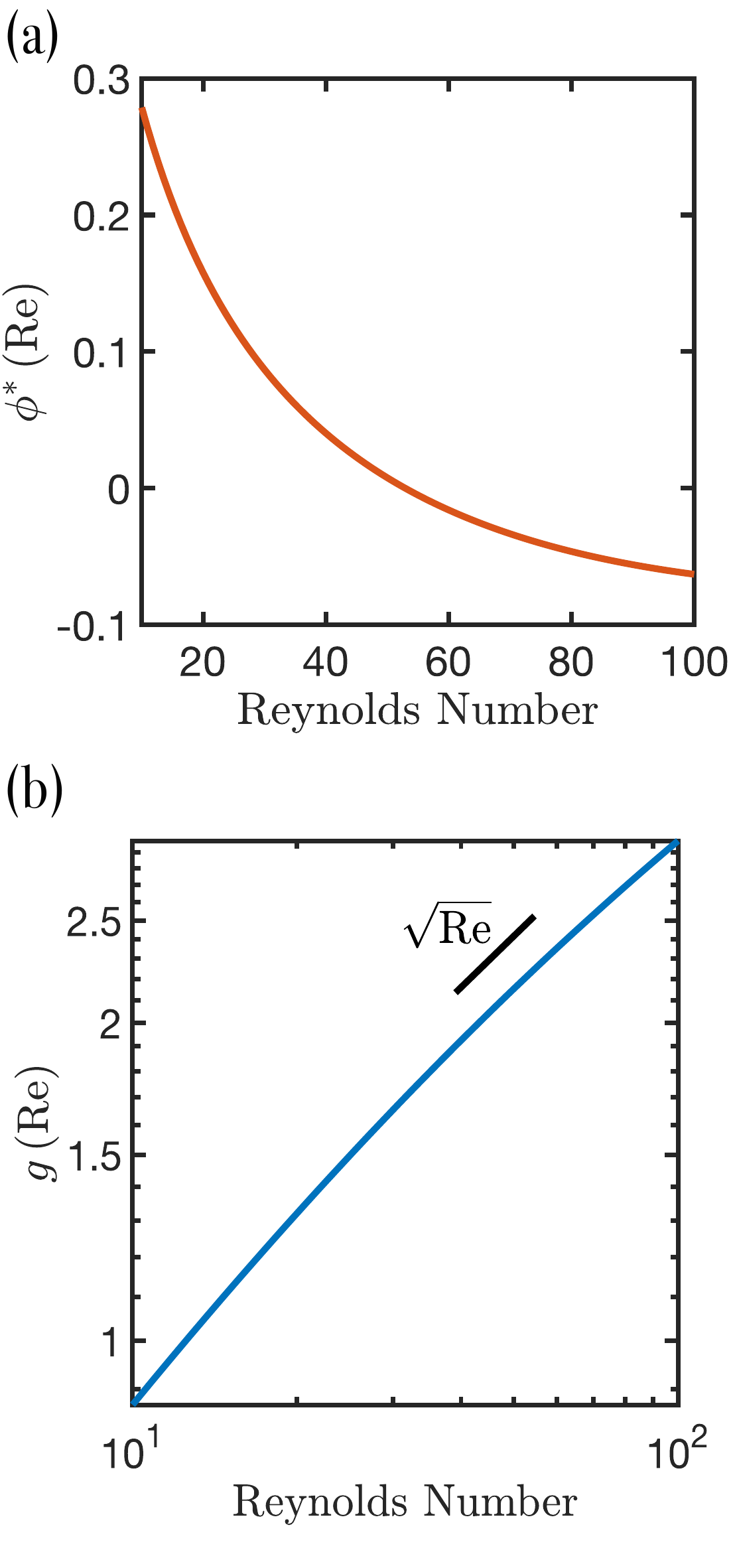}
    \caption{The optimal phase (a) $\phi^*(\text{Re})$ and maximum of the net force (scaled by $\epsilon^{-3}$) on the cylinder in the annulus (b) $g(\text{Re})$ as a function of the Reynolds number.} 
\label{Fig:Net Force Reynolds Number Function}
\end{figure}

In Figure \ref{Fig:Net Force Reynolds Number}(b), we plot the net force as a function of phase $-\pi \leq \phi \leq \pi$, and we again observe that the force appears sinusoidal. The amplitude and phase at which the maximal force occurs both change with Reynolds number.  We fit the function $\cos(\phi - \phi^*)$ to the normalized data for each Reynolds number to obtain how the optimal phase changes with Reynolds number. 
We plot the function $\phi^*(\text{Re})$ in Figure \ref{Fig:Net Force Reynolds Number Function}(a), and the optimal phase decreases from about 0.16 at Reynolds number 20 to about -0.06 at Reynolds number 100.
For this range, the force generated with $\phi=0$ is less then 2\% lower than the force generated with $\phi=\phi^{*}$.

We let $g(\text{Re})$ denote the maximum force as a function of the Reynolds number, which is found from computing the force at $\beta=1/\sqrt{2}$ and $\phi=\phi^{*}(\text{Re})$. Figure \ref{Fig:Net Force Reynolds Number Function}(b) shows that the size of the force scales at a rate little more than $\sqrt{\text{Re}}$. Finally, from this analysis we identified that the net force can be expressed as 
\begin{equation}
\text{F}_{net} = \epsilon^3f(\beta)g(\text{Re})\cos\left(\phi - \phi^*(\text{Re})\right),
\label{Eq:Dividing Up Net Force}
\end{equation}
where $f(\beta)$ is defined analytically in \eqref{Eq:Value of Function f}, and the optimal phase $\phi^{*}(\text{Re})$ and amplitude $g(\text{Re})$ are plotted in Figure \ref{Fig:Net Force Reynolds Number Function}. In Appendix \ref{Sec:App6}, we consider other domain sizes and show that the optimal phase, $\phi^{*}(\text{Re})$, is qualitatively similar on different domain sizes. The force amplitude, $g(\text{Re})$, exhibits the same approximate $\sqrt{\text{Re}}$ scaling on different domains, but the size of the force decreases with increasing domain size.

In the expression for $\text{F}_{net}$, the amplitude dependence enters through $\epsilon^3f(\beta)$, which in terms of physical parameters is
\begin{equation}
    \epsilon^3f(\beta) = \frac{3\sqrt{3}}{2} \frac{A_{1}^2A_{2}}{R_{\text{in}}}.
\end{equation}
The optimal amplitude ratio $\beta=1/\sqrt{2}$ arises from maximizing this expression with $\epsilon = (A_1^2 + A_2^2)^{1/2}/R_{\text{in}}$ fixed.  If this quantity were optimized with a different constraint, a different optimal amplitude ratio would result.  
\section{Pumping Through a Channel}\label{Sec:Pumping and Net Force}
In the previous section, we used asymptotic analysis and numerical computation to derive the form of the net force on a cylinder from the streaming flow generated from two-frequency oscillations inside an annular domain.
While there is no pumping inside an annulus, we show in this section that our analysis on the net force can be applied to understand the flow through a channel. 
Specifically, we set up numerical simulations of a cylinder oscillating with two frequencies inside a channel and measure the net-directed flow by computing the flux. We evaluate the flux for select parameters and show that the flux is proportional to the net force generated in the annular domain.  Therefore, the results from analysis of the net force from Section \ref{Sec:Computational Results} are directly applicable to pumping.

\subsection{Measuring the Flux}
We use numerical simulations to compute the flow inside a channel of height 8 and length 32 (length scaled by the cylinder radius) with no-slip boundary conditions on top and bottom and periodic conditions in the horizontal direction. The horizontal position of the cylinder's center is given by equation \eqref{Eq:NonDim Displacement}. We solve for 250 periods, which is sufficiently long to allow the initial transient to decay. Details of the numerical methodology are provided in Appendix \ref{Sec:App5}.
The flux through a channel of height $H$ is defined as 
$$Q(t) = \int_{0}^Hu(x, y, t)\;dy,$$
 and we quantify pumping by the time-averaged flux over the period $T$:
 \begin{equation}
 \langle Q(t)\rangle = \frac{1}{T}\int_{t - T}^{t}Q(\tau)\;d\tau. \label{Eq:Time Average Flux}
 \end{equation}

\begin{figure}
    \centering
    \includegraphics[width=1\linewidth]{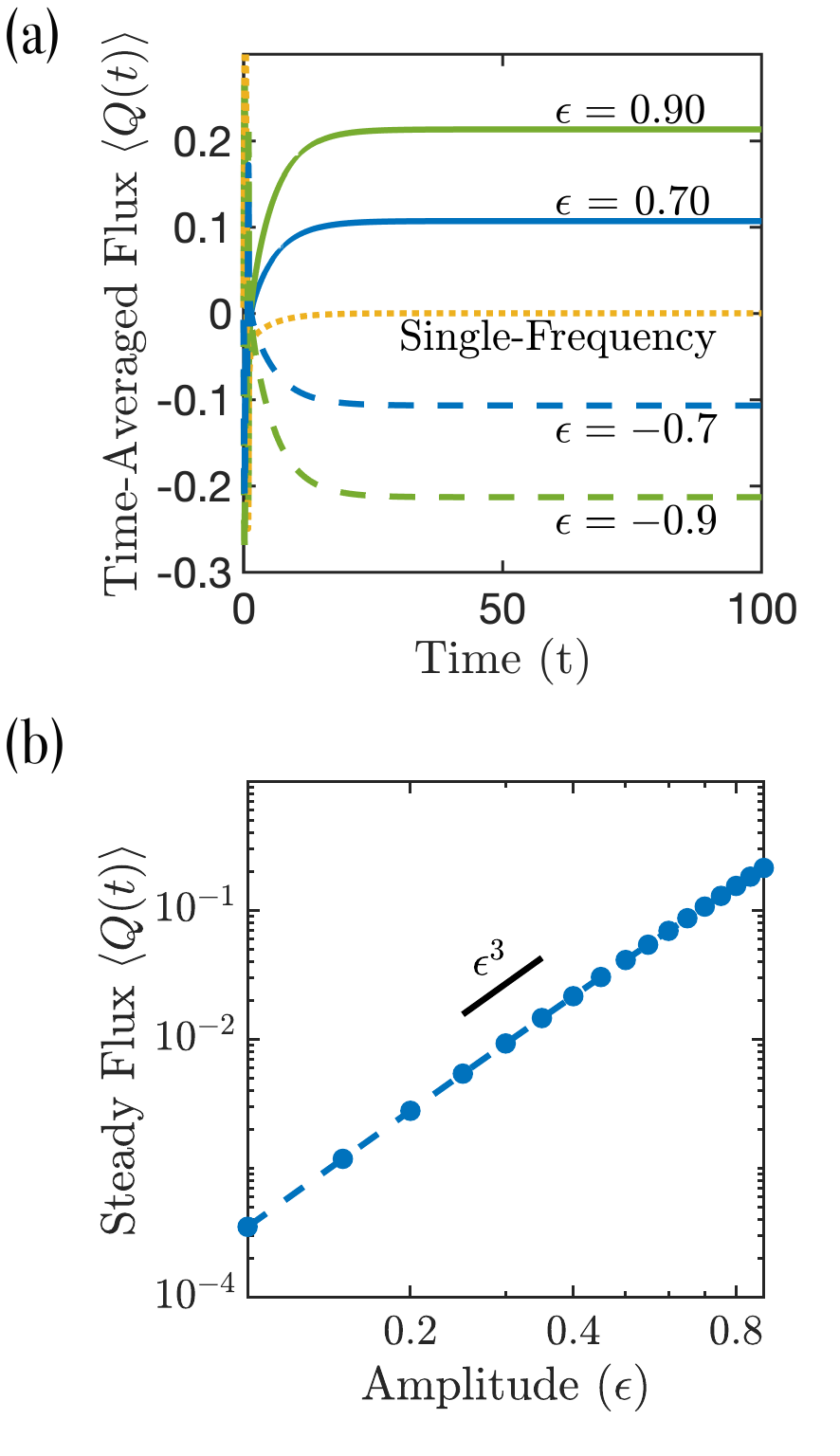}
    \caption{(a) The time-averaged flux through a rectangular channel over time for Re = 10, amplitude ratio $\beta = 1$, and phase $\phi = 0$ for different amplitudes $\epsilon$.
    (b) The steady time-averaged flux is third order in amplitude.}
\label{Fig:Measuring the Flux}
\end{figure}

Figure \ref{Fig:Measuring the Flux}(a) shows the period-averaged flux for amplitude ratio $\beta = 1$, phase $\phi = 0$, and Re = 10 as a function of time for different amplitude $\epsilon$, and the fluxes approach a steady value that depends on amplitude. Negating the amplitude results in negating the flux.  Following the same type of analysis from Section \ref{Sec:Computational Results Reversal}, the waveform generated by negative amplitude is equivalent to that obtained by shifting the phase difference $\phi$ by $\pi$. In the asymptotic analysis, the terms that generate a nonzero net force occur at third order in amplitude. Figure \ref{Fig:Measuring the Flux}(b) shows that the flux also exhibits third order scaling with amplitude. 

\begin{figure*}
    \centering
    \includegraphics[width=0.9\linewidth]{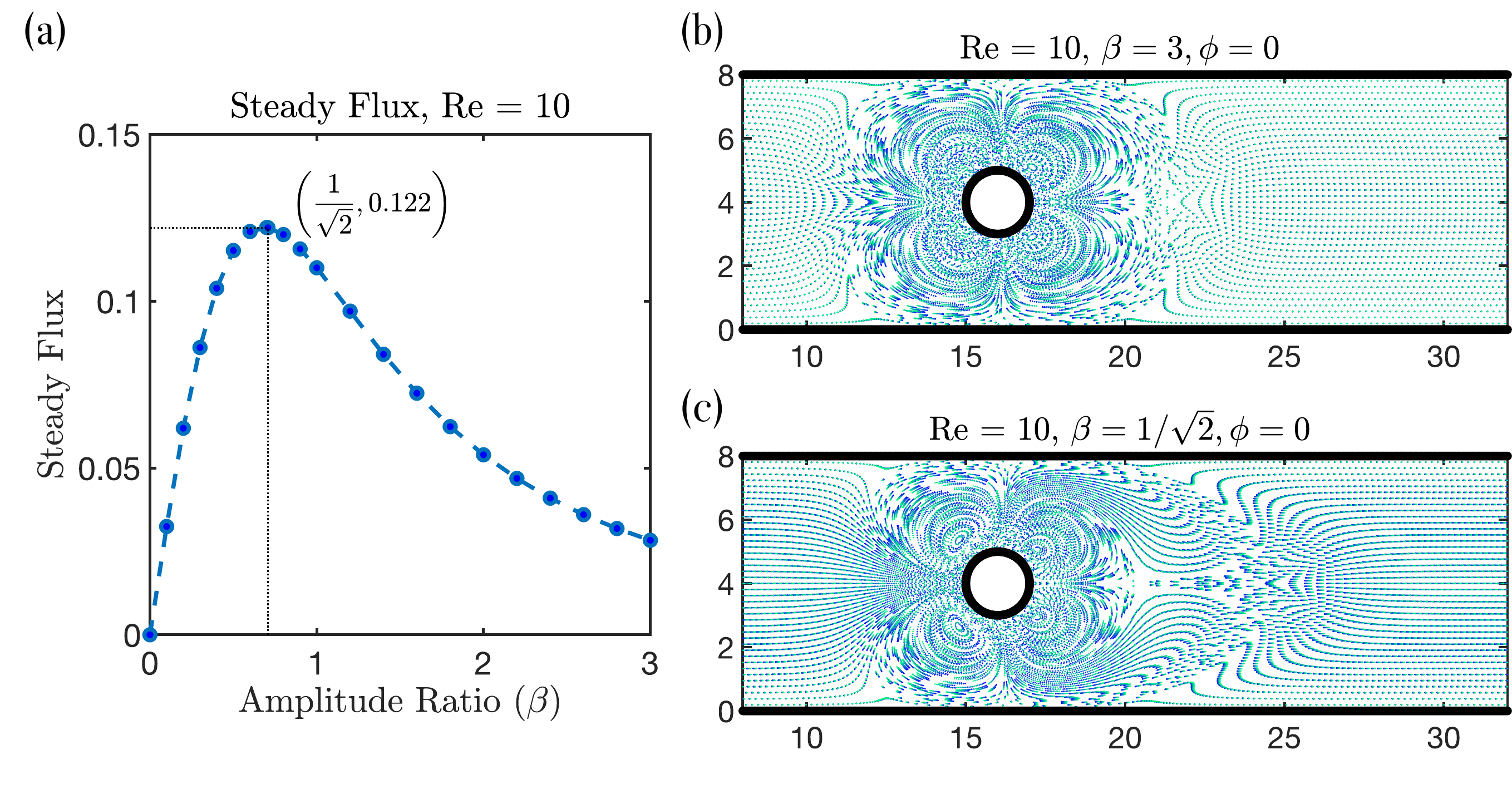}
    \caption{(a) The time-averaged flux for different amplitude ratios. The maximum flux is achieved at approximately $\beta \approx 1/\sqrt{2}$. Visualization of streaming flows in a 32 by 8 channel at time $T = 250$ with phase $\phi = 0$, amplitude $\epsilon = 0.5$, and Re = 10 for amplitude ratio (b) $\beta = 3$ and (c) $\beta = 1/\sqrt{2}$. Passive tracer particles highlight the positions of the flow from the last $10$ periods, where the current location is colored green and the location $10$ periods prior is colored blue.}
    \label{Fig:Pumping Visuals}
\end{figure*}

Our results in Section \ref{Sec:Computational Results} suggest that a cylinder oscillating with two frequencies will generate the largest pumping at amplitude ratio $\beta = 1/\sqrt{2}$. In Figure \ref{Fig:Pumping Visuals}(a), we show the flux computed at $\phi = 0$, $\epsilon = 0.5$, and Re = 10 for different amplitude ratios, and the flux appears to be maximized at  approximately $\beta \approx 1/\sqrt{2}$. The curve through these points resembles the graph of $f(\beta)$ from \eqref{Eq:Value of Function f} which describes how the net force in the annulus changes with amplitude ratio.  In the following section, we analyze the scaling between flux and force for different parameters. 

We show the flow field at $\beta = 1/\sqrt{2}$ in Figure \ref{Fig:Pumping Visuals}(b) at $\beta = 3$ in Figure \ref{Fig:Pumping Visuals}(c) to examine how the flow pattern changes with amplitude ratio. In both cases the net flow is left-to-right \--- away from the cylinder, the green particles (most current period) are to the right of the blue particles (past period), indicating a net rightward flow. However, the flow for amplitude ratio $\beta = 1/\sqrt{2}$ results in a larger amount of tracer particles moving to the right, consistent with stronger pumping.

\subsection{Comparing Flux and Net Force}
\begin{figure*}
    \centering
    \includegraphics[width=1\linewidth]{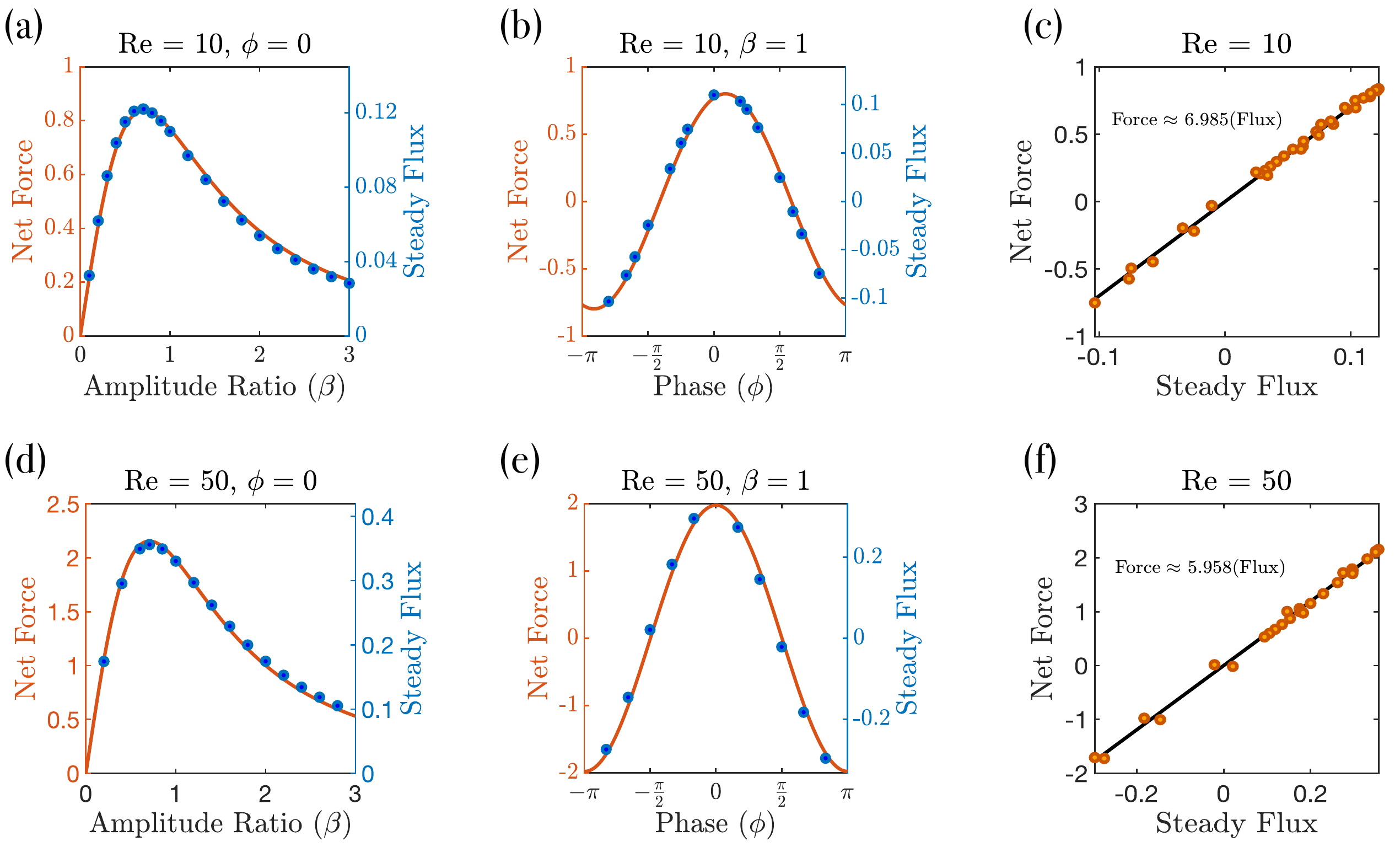}
    \caption{Net force on the cylinder computed on the annular domain (left axes) and steady time-averaged flux (right axes) through a channel for $\epsilon = 0.5$ for (a,b) Re = 10 and (d,e) Re = 50 as a function of (a,d) amplitude ratio and (b,e) phase. The net force versus the steady time-averaged flux at (c) Re = 10 and (f) Re = 50 are proportional to one another.  The solid line is the best fit line through the origin.} 
\label{Fig:Beta and Phi Comparison}
\end{figure*}

We compare the net force on the cylinder in the annular domain from our analysis with the time-averaged flux through a channel from our simulations. Figure \ref{Fig:Beta and Phi Comparison} shows the net force and time-averaged flux on the same plots as a function of amplitude ratio for fixed phase and as a function of phase for fixed amplitude ratio for two different Reynolds numbers. That flux data from the simulations (dots) show the same response to changes in parameters as the net force computed from the analysis (solid lines), and these plots suggest the flux and force are proportional to one another. 

For each numerical simulation, we compute the net force from the analysis for the same amplitude, phase and Reynolds number, and we scatter plot the net force against the steady flux for each Reynolds number in Figure \ref{Fig:Beta and Phi Comparison}(c) and (f). These plots show the flux is proportional to the net force with a proportionality constant that depends on the Reynolds number.  Therefore the results of the analysis on the net force in the annulus can be used to understand how the strength of pumping varies with parameters.

\section{Discussion}
We use asymptotic analysis and numerical simulations to analyze the pumping of a cylinder oscillating with two frequencies of different amplitudes and phases.
We analyze the flow inside a fixed annular domain with a prescribed oscillatory horizontal velocity of \eqref{Eq:NonDim Far Field} on the outer cylinder. 
While there is no pumping inside an annulus, we compute the net force and show that it has the form \eqref{Eq:Dividing Up Net Force}, which separates the effects of changes in amplitude, phase, and Reynolds number. For fixed amplitude, the net force achieves its maximal value at $\beta = 1/\sqrt{2}$, regardless of the phase difference and Reynolds number, and the optimal phase depends on the Reynolds number. The force can be reversed by changing the phase difference by $\pi$. By comparing with numerical simulations in a channel, we find that the net force inside an annular domain is proportional to the time-averaged flux through the channel. Thus our analysis of the force in the annulus can be used to  understand how pumping varies with different parameters. 

In our analysis of the flow between two concentric cylinders, we prescribe a two-frequency velocity at the outer wall and fix frequency ratio as $\alpha = 2$. 
Lee \textit{et al.}\cite{lee2025pumping} showed that the largest net force occurs at third-order in amplitude for frequency ratio $\alpha = 2$.
Other frequency ratios result in a force at higher order, with the next largest being fifth order. 
Our analysis can be adapted to general frequency ratios; however, we would need to compute the solution structures to at least fifth order and numerically solve an additional number of boundary value problems. 

The analysis and numerical simulations of the flow around a cylinder oscillating with two frequencies from Lee \textit{et al.}\cite{lee2025pumping} were inspired by experiments of an object sliding on a surface undergoing two-frequency lateral oscillations in Hashemi \textit{et al.}\cite{hashemi2022net}. The results from Lee \textit{et al.}\cite{lee2025pumping} gave necessary conditions on the frequency ratio for pumping that are identical to those of Hashemi \textit{et al.}\cite{hashemi2022net} However, our prior analysis did not elucidate how the pumping rate varies with different parameters, and thus, in this paper, we examined how the flow changes with amplitude ratio, phase, and Reynolds number.
Similarly, Zhang \textit{et al.}\cite{zhang2024theoretical} analytically and Hui \textit{et al.}\cite{hui2024vibrational} experimentally studied the motion of an object sliding on a surface undergoing two-frequency lateral oscillation for different amplitude ratios and the phase differences. 
Zhang \textit{et al.}\cite{zhang2024theoretical} showed that the largest net motion occurred at an amplitude ratio $\beta=1/4$ and at phase difference zero, independent of the friction coefficient. While a different physical system, we showed that the amplitude ratio $\beta = 1/\sqrt{2}$, corresponding to the largest pumping, is independent of Reynolds number, but the optimal phase difference changes with the Reynolds numbers. 

Single-frequency steady streaming has been exploited in microfluidic pumping \cite{marmottant2004bubble, tovar2009lateral, tovar2011lateral, huang2014reliable, zhang2024elasto}, but they require a shape asymmetry in the design of the device to produce a net-directed flow. 
For example, Tovar \& Lee \cite{tovar2009lateral} designed a micropump in which cavities of trapped air bubbles were at an angle along the channel walls. The angling of the cavities produces an asymmetric streaming flow that generates a net-directed flow through the channel.
We showed that a circular cylinder oscillating with two frequencies pumps fluid, and the direction and the strength of pumping can be controlled by the amplitude and phase of the two oscillations. The asymmetry in our system is temporal rather than spatial, and thus, fluid can be pumped in either direction without a built-in asymmetry in the device. Our results could lead to new designs of microfluidic pumps or potentially enhance the performance of existing designs.

\appendix
\section{Evaluating the Right-Hand Sides of \eqref{Eq:Bk}, \eqref{Eq:Ck}, and \eqref{Eq:Dk}} \label{Sec:App1}
To begin, we substitute $\psi_1$ from \eqref{Eq:First Order Solution} into the right-hand side of Equation \eqref{Eq:Second Order Line 1}, which is a quadratic nonlinearity, and the structure of the right-hand side is defined as 
\begin{eqnarray}
\text{Re}\; \boldsymbol{u_1}\boldsymbol{\cdot\nabla}\left(\Delta\psi_1\right) &=&  \Re\left(\sum_{k=1}^{4}\Omega_k(r)e^{-ikt}\right)\sin(2\theta)\nonumber\\
&& + \Omega_0(r)\sin(2\theta)\label{Eq:Second Order Right Hand Side}.
\end{eqnarray}
We simplify our notation of $\Omega_k(r)$ by defining the following oporators,
\begin{eqnarray}
Q_1[f(r), g(r), \ell_1, \ell_2] &=& \frac{\ell_1}{r}\left(f\cdot\frac{d^3g}{dr^3}\right) + \frac{\ell_1}{r^2} \left(f\cdot\frac{d^2g}{dr^2}\right)\nonumber\\
&& - \frac{\ell_1}{r^3} \left(f \cdot \frac{dg}{dr}\right) - \frac{\ell_1\ell_2^2}{r^3}\left(f\cdot \frac{dg}{dr}\right)\nonumber\\
&& + \frac{2\ell_1\ell_2^2}{r^4}\left(f \cdot g\right), \label{Eq:Right-Hand Operator 1}
\end{eqnarray}
and 
\begin{eqnarray}
Q_2[f(r), g(r), \ell_2] &=& -\frac{\ell_2}{r}\left(\frac{df}{dr}\cdot \frac{d^2g}{dr^2}\right) - \frac{\ell_2}{r^2}\left(\frac{df}{dr} \cdot \frac{dg}{dr}\right)\nonumber\\
&&+ \frac{\ell_2^3}{r^3}\left(\frac{df}{dr} \cdot g\right).\label{Eq:Right-Hand Operator 2}
\end{eqnarray}
We use the operators \eqref{Eq:Right-Hand Operator 1} and \eqref{Eq:Right-Hand Operator 2} to define $\Omega_k$ as, 
\begin{eqnarray*}
\Omega_0 &=& \frac{\text{Re}}{8}\left(Q_1[a_1^*(r), a_1(r), 1, 1] + Q_1[a_1(r), a_1^*(r), 1, 1]\right.\\
&&+ Q_1[a_2^*(r), a_2(r), 1, 1] + Q_1[a_2(r), a_2^*(r), 1, 1]\\
&&+ Q_2[a_1^*(r), a_1(r), 1] + Q_2[a_1(r), a_1^*(r), 1]\\
&&\left. + Q_2[a_2^*(r), a_2(r), 1] + Q_2[a_2(r), a_2^*(r), 1]\right),
\end{eqnarray*}
\begin{eqnarray*}
\Omega_1 &=& \frac{\text{Re}}{4}\left(Q_1[a_1^*(r), a_2(r), 1, 1] + Q_1[a_2(r), a_1^*(r), 1, 1]\right.\\
&&\left. + Q_2[a_1^*(r), a_2(r), 1] + Q_2[a_2(r), a_1^*(r), 1]\right),
\end{eqnarray*}
\begin{eqnarray*}
\Omega_2 &=&\frac{\text{Re}}{4}\left(Q_1[a_1(r), a_1(r), 1, 1] + Q_2[a_1(r), a_1(r), 1]\right),
\end{eqnarray*}
\begin{eqnarray*}
\Omega_3 &=& \frac{\text{Re}}{4}\left(Q_1[a_1(r), a_2(r), 1, 1] + Q_1[a_2(r), a_1(r), 1, 1]\right.\\
&&\left. + Q_2[a_1(r), a_2(r), 1] + Q_2[a_2(r), a_1(r), 1]\right),
\end{eqnarray*}
\begin{eqnarray*}
\Omega_4 &=&\frac{\text{Re}}{4}\left(Q_1[a_2(r), a_2(r), 1, 1] + Q_2[a_2(r), a_2(r), 1]\right),
\end{eqnarray*}
where $a_1(r)$ and $a_2(r)$ are the solutions of \eqref{Eq:Ak}-\eqref{Eq:Ak Bd 3}.
The $^*$ denotes the complex conjugate.

Similarly, we substitute $\psi_1$ and $\psi_2$ from \eqref{Eq:First Order Solution} and \eqref{Eq:Second Order Solution} into the right-hand side of \eqref{Eq:Third Order Line 1} and obtain its structure as, 
\begin{gather}
\text{Re}\;\boldsymbol{u_2}\boldsymbol{\cdot\nabla}\left(\Delta\psi_1\right) + \text{Re}\; \boldsymbol{u_1}\boldsymbol{\cdot\nabla}\left(\Delta\psi_2\right)\nonumber\\
=\Re\left(\sum_{k=1}^{6}\rho_k(r)e^{-ikt}\right)\sin(3\theta) + \rho_0(r)\sin(3\theta)\nonumber\\
+\Re\left(\sum_{k=1}^{6}\xi_k(r)e^{-ikt}\right)\sin(\theta) + \xi_0(r)\sin(\theta). \label{Eq:Third Order Right Hand Side}
\end{gather}
It should be noted that we will only solve for $\rho_k(r)$ and $\xi_k(r)$ from \eqref{Eq:Third Order Right Hand Side} at $k = 0$, as $\rho_0(r)$ and $\xi_0(r)$ are the terms in the right-hand side that lead to a steady solution at third order. 
In \eqref{Eq:Ck}-\eqref{Eq:Dk Bd 2}, we solve for $d_0(r)$ to compute the net force and solve for $c_0(r)$ as well to generate the contours of the steady flow.  
Therefore, we will omit the derivation of $\rho_k(r)$ and $\xi_k(r)$ for other $k$ values.
In general, we define 
$$\xi_k(r) = \tau_k^{(1)} - \tau_k^{(2)}, \quad \rho_k(r) = \tau_k^{(1)} + \tau_k^{(2)},$$
where 
\begin{eqnarray}
\tau_0^{(1)} &=& \frac{\text{Re}}{8} \left(Q_1\left[a_1^*(r), b_1(r), 1, 2\right] + Q_2[b_1(r), a_1^*(r), 1]\right.\nonumber\\
&+&Q_1[a_1(r), b_1^*(r),1, 2] + Q_2[b_1^*(r), a_1(r), 1]\nonumber\\ 
&+&Q_1[a_2^*(r), b_2(r),1, 2] + Q_2[b_2(r), a_2^*(r), 1]\nonumber\\
&+& \left.Q_1[a_2(r), b_2^*(r),1, 2]+Q_2[b_2^*(r), a_2(r), 1]\right),\nonumber
\end{eqnarray}
and
\begin{eqnarray}
\tau_0^{(2)} &=& \frac{\text{Re}}{8}\left(Q_2[a_2^*(r), b_2(r), 2] + Q_1[b_2(r), a_2^*(r), 2, 1]\right.\nonumber\\
&+& Q_2[a_2(r), b_2^*(r), 2] + Q_1[b_2^*(r), a_2(r), 2, 1]\nonumber\\
&+& Q_2[a_1(r), b_1^*(r), 2]+Q_1[b_1^*(r), a_1(r), 2, 1]\nonumber\\
&+& \left. Q_2[a_1^*(r), b_1(r), 2]+Q_1[b_1(r), a_1^*(r), 2, 1]\right).\nonumber
\end{eqnarray}
$a_k(r)$ and $b_k(r)$ are the solutions of \eqref{Eq:Ak}-\eqref{Eq:Ak Bd 3} and \eqref{Eq:Bk}-\eqref{Eq:Bk Bd 2}, correspondingly. 

\section{Equations for Pressure} \label{Sec:App2}
In this section, we obtain the equations at third order that involve a steady pressure, which contributes a net force. 
To compute the pressure $p$, we convert our stream function, 
$$\psi = \epsilon\psi_1 +  \epsilon^2\psi_2 +  \epsilon^3\psi_3 + O\left(\epsilon^4\right)$$
into velocity components, 
\begin{gather}
u_r = \epsilon u_{r, 1} + \epsilon^2 u_{r, 2} + \epsilon^3 u_{r, 3} + O\left(\epsilon^4\right),\label{Eq:Radial Velocity Expansion}\\
u_\theta = \epsilon u_{\theta, 1} + \epsilon^2 u_{\theta, 2} + \epsilon^3 u_{\theta, 3} + O\left(\epsilon^4\right),\label{Eq:Theta Velocity Expansion}
\end{gather}
where 
$$u_{r,n} = \frac{1}{r}\frac{\partial \psi_n}{\partial \theta}, \quad u_{\theta,n} = -\frac{\partial \psi_n}{\partial r}.$$
We expand the pressure in the limit of small amplitude,
\begin{equation}
p = \epsilon p_1 +  \epsilon^2 p_2 +  \epsilon^3 p_3 + O\left(\epsilon^4\right), \label{Eq:Pressure Expansion}
\end{equation}
and substitute the $u_r$, $u_\theta$, and $p$ of Equations \eqref{Eq:Radial Velocity Expansion}-\eqref{Eq:Pressure Expansion} into the momentum equation,
\begin{eqnarray}
\frac{\partial p}{\partial r} &=& \Delta u_r - \frac{u_r}{r^2} - \frac{2}{r^2}\frac{\partial u_\theta}{\partial \theta}\nonumber\\
&&-\text{Re}\left(\frac{\partial u_r}{\partial t} + u_r\frac{\partial u_r}{\partial r} + \frac{u_\theta}{r}\frac{\partial u_r}{\partial \theta} - \frac{u_\theta^2}{r}\right),\label{Pressure Derivative Equation}
\end{eqnarray}
\begin{eqnarray}
\frac{1}{r}\frac{\partial p}{\partial \theta} &=& \Delta u_\theta - \frac{u_\theta}{r^2} + \frac{2}{r^2}\frac{\partial u_r}{\partial \theta}\nonumber\\
&&-\text{Re}\left(\frac{\partial u_\theta}{\partial t} + u_r\frac{\partial u_\theta}{\partial r} + \frac{u_\theta}{r}\frac{\partial u_\theta}{\partial \theta} + \frac{u_r u_\theta}{r}\right).\label{Pressure Equation}
\end{eqnarray}
Consequently, we obtain successive equations at each order:
\begin{gather}
\textbf{First Order}\nonumber\\
\label{Eq:Pressure First Order Line 1}
\frac{\partial p_1}{\partial r} = \Delta u_{r,1} - \frac{u_{r,1}}{r^2} - \frac{2}{r^2}\frac{\partial u_{\theta,1}}{\partial \theta} -\text{Re}\frac{\partial u_{r,1}}{\partial t},
\end{gather}
\begin{gather}
\label{Eq:Pressure First Order Line 2}
\frac{1}{r}\frac{\partial p_1}{\partial \theta} = \Delta u_{\theta,1} - \frac{u_{\theta,1}}{r^2} + \frac{2}{r^2}\frac{\partial u_{r,1}}{\partial \theta} -\text{Re}\frac{\partial u_{\theta,1}}{\partial t},
\end{gather}
\begin{gather}
\textbf{Second Order}\nonumber\\
\frac{\partial p_2}{\partial r} = \Delta u_{r,2} - \frac{u_{r,2}}{r^2} - \frac{2}{r^2}\frac{\partial u_{\theta,2}}{\partial \theta} -\text{Re}\frac{\partial u_{r,2}}{\partial t}\nonumber\\
-\text{Re}\left(u_{r,1}\frac{\partial u_{r,1}}{\partial r} + \frac{u_{\theta,1}}{r}\frac{\partial u_{r, 1}}{\partial \theta} - \frac{u_{\theta, 1}^2}{r}\right),\label{Eq:Pressure Second Order Line 1}
\end{gather}
\begin{gather}
\frac{1}{r}\frac{\partial p_2}{\partial \theta} = \Delta u_{\theta,2} - \frac{u_{\theta,2}}{r^2} + \frac{2}{r^2}\frac{\partial u_{r,2}}{\partial \theta} -\text{Re}\frac{\partial u_{\theta,2}}{\partial t},\nonumber\\
-\text{Re}\left(u_{r,1}\frac{\partial u_{\theta,1}}{\partial r} + \frac{u_{\theta,1}}{r}\frac{\partial u_{\theta,1}}{\partial \theta} + \frac{u_{r,1} u_{\theta,1}}{r}\right),\label{Eq:Pressure Second Order Line 2}
\end{gather}
\begin{gather}
\textbf{Third Order}\nonumber\\
\frac{\partial p_3}{\partial r} = \Delta u_{r,3} - \frac{u_{r,3}}{r^2} - \frac{2}{r^2}\frac{\partial u_{\theta,3}}{\partial \theta} -\text{Re}\frac{\partial u_{r,3}}{\partial t}\nonumber\\
-\text{Re}\left(u_{r,2}\frac{\partial u_{r,1}}{\partial r} + \frac{u_{\theta,2}}{r}\frac{\partial u_{r, 1}}{\partial \theta} - \frac{u_{\theta, 2}u_{\theta, 1}}{r}\right)\nonumber\\
\qquad-\text{Re}\left(u_{r,1}\frac{\partial u_{r,2}}{\partial r} + \frac{u_{\theta,1}}{r}\frac{\partial u_{r, 2}}{\partial \theta} - \frac{u_{\theta, 1}u_{\theta, 2}}{r}\right),\label{Eq:Pressure Third Order Line 1}
\end{gather}
\begin{gather}
\frac{1}{r}\frac{\partial p_3}{\partial \theta} = \Delta u_{\theta,3} - \frac{u_{\theta,3}}{r^2} + \frac{2}{r^2}\frac{\partial u_{r,3}}{\partial \theta} -\text{Re}\frac{\partial u_{\theta,3}}{\partial t}\nonumber\\
-\text{Re}\left(u_{r,2}\frac{\partial u_{\theta,1}}{\partial r} + \frac{u_{\theta,2}}{r}\frac{\partial u_{\theta,1}}{\partial \theta} + \frac{u_{r,2} u_{\theta,1}}{r}\right)\nonumber\\
\qquad-\text{Re}\left(u_{r,1}\frac{\partial u_{\theta,2}}{\partial r} + \frac{u_{\theta,1}}{r}\frac{\partial u_{\theta,2}}{\partial \theta} + \frac{u_{r,1} u_{\theta,2}}{r}\right).\label{Eq:Pressure Third Order Line 2}
\end{gather}

We convert $\psi_1$ from \eqref{Eq:First Order Solution} as velocity components, $u_{r,1}$ and $u_{\theta,1}$, and substitute the values into the right-hand side of \eqref{Eq:Pressure First Order Line 1} and \eqref{Eq:Pressure First Order Line 2} to obtain the solution structure of $p_1$ as
\begin{equation}
\label{Eq:Pressure First Order Solution}
p_1 = \Re\left(A_1(r)e^{-it} + A_2(r)e^{-i2t}\right)\cos(\theta).
\end{equation}
Similarly, we use both $\psi_1$ and $\psi_2$ from \eqref{Eq:First Order Solution} and \eqref{Eq:Second Order Solution} to obtain the right-hand side of \eqref{Eq:Pressure Second Order Line 1} and \eqref{Eq:Pressure Second Order Line 2}.
We define the form of the solution at second order as, 
\begin{gather}
p_2 = \left(\sum_{k=1}^4 B_k(r)e^{-ikt}\right)\cos(2\theta) + B_0(r)\cos(2\theta)\nonumber\\
+\left(\sum_{k=1}^4 E_k(r)e^{-ikt}\right) + E_0(r).\label{Eq:Pressure Second Order Solution}
\end{gather}

Finally, we use $\psi_1$, $\psi_2$, and $\psi_3$ from \eqref{Eq:First Order Solution}-\eqref{Eq:Third Order Solution} to obtain the right-hand side of \eqref{Eq:Pressure Third Order Line 1} and \eqref{Eq:Pressure Third Order Line 2} and derive the solution structure of $p_3$ as,
\begin{gather}
p_3 = \left(\sum_{k=1}^6 C_k(r)e^{-ikt}\right)\cos(3\theta) + C_0(r)\cos(3\theta)\nonumber\\
+\left(\sum_{k=1}^6 D_k(r)e^{-ikt}\right)\cos(\theta) + D_0(r)\cos(\theta).\label{Eq:Pressure Third Order Solution}
\end{gather}
From the net force computation of \eqref{Eq:Net Force}, $D_0(r)$ is the only term in $p_3$ that contributes a net force.  
From the right-hand side of \eqref{Eq:Pressure Third Order Line 1} and \eqref{Eq:Pressure Third Order Line 2}, $D_0(r)$ is involved in a PDE that linearly depends on $\psi_3$ and contains a quadratic nonlinearity of $\psi_1$ and $\psi_2$. 
Specifically, we define a linear operators of the form, 
\begin{gather}
L_1\left[f\right] = r \frac{d^3f}{dr^3} + \frac{d^2f}{dr^2} -\frac{2}{r}\; \frac{df}{dr} + \frac{2}{r^2}\;f, \label{Eq:Pressure Linear Operator 1}
\end{gather}
and 
\begin{gather}
L_2\left[f\right] = \frac{1}{r}\frac{d^2f}{dr^2} + \frac{1}{r^2}\frac{df}{dr} - \frac{1}{r^3}\;f. \label{Eq:Pressure Linear Operator 2}
\end{gather}
We also define the nonlinear operators as
\begin{gather}
NL_1^{m, n}\left[f, g\right] = \text{sgn}(n-m) \left(\frac{m}{2}\;f\;\frac{d^2g}{dr^2}\right.\nonumber\\ 
+ \left.\frac{n}{2}\;\frac{df}{dr}\;\frac{dg}{dr} + \frac{m}{2r}\;f\;\frac{dg}{dr}\right), \label{Eq:Pressure NonLinear Operator 1}
\end{gather}
and 
\begin{gather}
NL_2^{m, n}\left[f, g\right] = \frac{mn}{2r^2}\;f\;\frac{dg}{dr} - \frac{mn}{2r^3}\;f\;g\nonumber\\
+ \frac{n^2}{2r^2}\;\frac{df}{dr}\;g - \frac{1}{2r}\;\frac{df}{dr}\frac{dg}{dr}. \label{Eq:Pressure NonLinear Operator 2}
\end{gather}
Substituting \eqref{Eq:Pressure Third Order Solution} and \eqref{Eq:First Order Solution}-\eqref{Eq:Third Order Solution} into \eqref{Eq:Pressure Third Order Line 2}, we obtain
\begin{equation}
D_0(r) = L_1\left[d_0\right] - \text{Re}\;NL^{(1)},\label{Eq:Pressure Solution}
\end{equation}
where 
\begin{gather*}
NL^{(1)} = \frac{1}{4}\left(NL_1^{2, 1}\left[b_2, a_2^*\right] + NL_1^{2, 1}\left[b_1, a_1^*\right]\right.\\
+ NL_1^{2, 1}\left[b_2^*, a_2\right] + NL_1^{2, 1}\left[b_1^*, a_1\right] + NL_1^{1, 2}\left[a_2, b_2^*\right]\\
+ \left.NL_1^{1, 2}\left[a_1, b_1^*\right] + NL_1^{1, 2}\left[a_2^*, b_2\right] + NL_1^{1, 2}\left[a_1^*, b_1\right]\right).
\end{gather*}


\section{Numerical Solutions of \eqref{Eq:Ak}-\eqref{Eq:Dk Bd 2}}\label{Sec:App3}
We compute the solutions of Equations \eqref{Eq:Ak}-\eqref{Eq:Dk Bd 2} using pseudo-spectral methods. 
The differentiation matrices are constructed by using Chebyshev polynomials, which are defined on $s \in [-1, 1]$, and in the computation of the net force, we use 120 Chebyshev points. 
We solve the problem on $r \in [1, R]$, and thus, we apply a linear transformation, 
\begin{equation}
\label{Eq:Linear Transformation}
r = 1 + \frac{(R-1)(s+1)}{2},
\end{equation}
which maps $s = -1$ and $s = 1$ to $r = 1$ and $r = R$, correspondingly.
Moreover, the derivative with respect $r$ is related to those of $s$ by  
\begin{equation}
\label{Eq:Derivative Transformation}
\frac{d}{dr} = \frac{2}{R - 1}\frac{d}{ds}.
\end{equation}
\subsection{First Order Computation}
In the following, we carry out the steps to compute solution $a_k(r)$ of Equations \eqref{Eq:Ak}-\eqref{Eq:Ak Bd 3} for $k = 1$ and $k = 2$.
Expanding the differential operator \eqref{Eq: Differential Operator} in Equation \eqref{Eq:Ak}, we obtain 
\begin{eqnarray}
&a_k^{(4)} + \frac{2}{r}\;a_k^{(3)} + \left(ik\;\text{Re} - \frac{3}{r^2}\right)a_k''\nonumber\\
& + \left(\frac{ik\;\text{Re}}{r} + \frac{3}{r^3}\right)a_k' + \left(-\frac{ik\;\text{Re}}{r^2} - \frac{3}{r^4}\right)a_k = 0.\label{Eq:First Order Numerical Equation 1}
\end{eqnarray}
Substituting the derivatives in Equation \eqref{Eq:First Order Numerical Equation 1} with derivatives \eqref{Eq:Derivative Transformation} that are dependent on $s\in[-1, 1]$, we obtain 
\begin{eqnarray}
&\frac{16}{(R - 1)^4}\;a_k^{(4)} + \frac{16}{(R - 1)^3r}a_k^{(3)} + \left(\frac{i4k\;\text{Re}}{(R - 1)^2} - \frac{12}{(R-1)^2r^2}\right)a_k''\nonumber\\
& + \left(\frac{i2k\;\text{Re}}{(R - 1)r} + \frac{6}{(R - 1)r^3}\right)a_k' + \left(-\frac{3}{r^4} - \frac{ik\;\text{Re}}{r^2}\right)a_k = 0.
\label{Eq:First Order Numerical Equation 2}
\end{eqnarray}
Because of the inhomogeneous boundary conditions at $r = R$ or $s = 1$ in \eqref{Eq:Ak Bd 2} and \eqref{Eq:Ak Bd 3}, we define a polynomial interpolant, $p_k(s) = a_k(s) - v_k(s)$ where $v_k(s)$ satisfies homogeneous boundary conditions.
To simplify the notation, we define the constants $c_1^{(k)}$ and $c_2^{(k)}$ where
$$c_1^{(1)} = -\frac{R}{\sqrt{1+\beta^2}}, \quad c_2^{(1)} = -\frac{1}{\sqrt{1+\beta^2}},$$
and
$$c_1^{(2)} = -\frac{R\; 2\beta e^{-i\phi}}{\sqrt{1+\beta^2}}, \quad c_2^{(2)} = -\frac{2\beta e^{-i\phi}}{\sqrt{1+\beta^2}},$$
which implies that $p_k(s)$ has the form,
\begin{eqnarray}
p_k(s) &=& \left(\frac{c_2^{(k)}}{4} - \frac{c_1^{(k)}}{4}\right)s^3 + \frac{c_2^{(k)}}{4}\;s^2\nonumber\\ 
&&+\left(\frac{3c_1^{(k)}}{4} - \frac{c_2^{(k)}}{4}\right)\;s + \left(\frac{c_1^{(k)}}{2} - \frac{c_2^{(k)}}{4}\right).
\label{Eq:First Order Numerical Right Hand}
\end{eqnarray}
Equation \eqref{Eq:First Order Numerical Equation 2} in terms of $v(s)$ results in
\begin{eqnarray}
&\frac{16}{(R - 1)^4}\;v_k^{(4)} + \frac{16}{(R - 1)^3r}v_k^{(3)} + \left(\frac{i4k\;\text{Re}}{(R - 1)^2} - \frac{12}{(R-1)^2r^2}\right)v_k''\nonumber\\
& + \left(\frac{i2k\;\text{Re}}{(R - 1)r} + \frac{6}{(R - 1)r^3}\right)v_k' + \left(-\frac{3}{r^4} - \frac{ik\;\text{Re}}{r^2}\right)v_k = \gamma_k,
\label{Eq:First Order Numerical Equation 3}
\end{eqnarray}
where $\gamma_k = \delta_1c_1^{(k)} + \delta_2c_2^{(k)}$ and 
\begin{eqnarray*}
\delta_1 &=& \left(-\frac{3}{4r^4}-\frac{ik\; \text{Re}}{4r^2}\right)s^3 + \left(\frac{9}{2r^3(R-1)} + \frac{i3k\; \text{Re}}{2r(R-1)}\right)s^2\\
&&+ \left(\frac{9}{4r^4} - \frac{18}{r^2(R-1)^2} + \frac{i6k\;\text{Re}}{(R-1)^2} + \frac{i3k\;\text{Re}}{4r^2}\right)s\\
&&-\frac{9}{2r^3(R-1)} + \frac{24}{r(R-1)^3} - \frac{i3k\;\text{Re}}{2r(R-1)}\\
&&+\frac{ik\;\text{Re}}{2r^2} + \frac{3}{2r^4},
\end{eqnarray*}
\begin{eqnarray*}
\delta_2 &=& \left(\frac{3}{4r^4} + \frac{ik\;\text{Re}}{4r^2}\right)s^3 + \left(\frac{3}{4r^4} - \frac{9}{2r^3(R-1)}\right.\\
&& \left.+ \frac{ik\;\text{Re}}{4r^2} - \frac{i3k\;\text{Re}}{2r(R-1)}\right)s^2\\
&& + \left(-\frac{3}{4r^4} - \frac{3}{r^3(R-1)} + \frac{18}{r^2(R-1)^2}\right.\\
&& \left.-\frac{ik\;\text{Re}}{4r^2} - \frac{i6k\;\text{Re}}{(R-1)^2} - \frac{ik\;\text{Re}}{r(R-1)}\right)s\\
&&+ \frac{3}{2r^3(R-1)} + \frac{6}{r^2(R-1)^2} - \frac{24}{r(R-1)^3} - \frac{ik\;\text{Re}}{4r^2}\\
&&-\frac{3}{4r^4} - \frac{i2k\;\text{Re}}{(R-1)^2} + \frac{ik\;\text{Re}}{2r(R-1)}.
\end{eqnarray*}
Finally, we define
\begin{equation*}
v_k(s) = (1 - s^2)q_k(s),
\end{equation*}
which limits the number of boundary conditions to two: 
\begin{equation*}
q_k(-1) = 0,\quad q_k(1) = 0,
\end{equation*}
and Equation \eqref{Eq:First Order Numerical Equation 3} becomes
\begin{equation*}
\sigma_4\; q_k^{(4)} + \sigma_3\; q_k^{(3)} + \sigma_2\; q''_k + \sigma_1\;q'_k + \sigma_0\;q_k= \gamma_k
\end{equation*}
where
\begin{eqnarray*}
\sigma_4 &=& \frac{16(1-s^2)}{(R-1)^4},
\end{eqnarray*}
\begin{eqnarray*}
\sigma_3 &=& -\frac{128s}{(R-1)^4} + \frac{16(1-s^2)}{r(R-1)^3},
\end{eqnarray*}
\begin{eqnarray*}
\sigma_2 &=& \left(\frac{12}{r^2(R-1)^2} - \frac{i4k\;\text{Re}}{(R-1)^2}\right)s^2 - \left(\frac{96s}{r(R-1)^3}\right)s\\
&&-\frac{192}{(R-1)^4}-\frac{12}{r^2(R-1)^2} + \frac{i4k\;\text{Re}}{(R-1)^2},\\
\end{eqnarray*}
\begin{eqnarray*}
\sigma_1 &=& \left(-\frac{6}{r^3(R-1)} - \frac{i2k\;\text{Re}}{r(R-1)}\right)s^2\\
&& + \left(\frac{48}{r^2(R-1)^2} - \frac{i16k\;\text{Re}}{(R-1)^2}\right)s\\
&& - \frac{96}{r(R-1)^3} + \frac{6}{r^3(R-1)} + \frac{i2k\;\text{Re}}{r(R-1)},\\
\end{eqnarray*}
\begin{eqnarray*}
\sigma_0 &=& \left(\frac{3}{r^4} + \frac{ik\;\text{Re}}{r^2}\right)s^2\\
&& + \left(-\frac{12}{r^3(R-1)} - \frac{i4k\;\text{Re}}{r(R-1)}\right)s\\
&& + \frac{24}{r^2(R-1)^2} - \frac{i8k\;\text{Re}}{(R-1)^2} - \frac{3}{r^4} - \frac{ik\; \text{Re}}{r^2}.
\end{eqnarray*}
Constructing the differentiation matrices with two homogeneous boundary conditions, we solve for $q_k$ by using MATLAB's backslash command and obtain $a_k$ by substituting $q_k$ into 
$$a_k(s) = p_k(s) + (1-s^2)q_k(s).$$
\subsection{Second and Third Order Solutions}
The computations required to solve $b_k(r), c_k(r),$ and $d_k(r)$ of Equations \eqref{Eq:Bk}-\eqref{Eq:Bk Bd 2}, \eqref{Eq:Ck}-\eqref{Eq:Ck Bd 2}, and \eqref{Eq:Dk}-\eqref{Eq:Dk Bd 2} are equivalent to each other. 
For simplicity, we solve for $f_{k, \ell}$, which is a solution of
\begin{gather}
  D_\ell\left(D_\ell + ik\;\text{Re}\right)\; f_{k,\ell}(r) = \text{RHS},\label{Eq:Fk}\\
  f_{k,\ell}(1) = 0, \quad  f_{k,\ell}'(1) = 0, \label{Eq:Fk Bd 1}\\
   f_{k,\ell}(R) = 0, \quad  f_{k,\ell}'(R) = 0.\label{Eq:Fk Bd 2}
\end{gather}
It should be noted that RHS of Equation \eqref{Eq:Fk} denotes the right-hand side of Equations \eqref{Eq:Bk}, \eqref{Eq:Ck}, and \eqref{Eq:Dk}, i.e., $\Omega_k(r)$, $\rho_k(r)$, and $\xi_k(r)$.  
We apply a linear transformation \eqref{Eq:Linear Transformation} that results Equation \eqref{Eq:Fk} to have derivatives of $s \in [-1, 1]$:
\begin{eqnarray}
&& a^{k, \ell}_4 f_{k, \ell}^{(4)}(s) + a^{k, \ell}_3 f_{k, \ell}^{(3)}(s) + a^{k, \ell}_2 f_{k, \ell}^{(2)}(s)\nonumber\\
&&+ a^{k, \ell}_1 f_{k, \ell}^{(1)}(s) + a^{k, \ell}_0 f_{k, \ell}^{(0)}(s) = \text{RHS}
\label{Eq:Second Order Numerical Equation 1}
\end{eqnarray}
where
$$a^{k, \ell}_4 = \frac{16}{(R-1)^4},$$
$$a^{k, \ell}_3 = \frac{16}{r(R-1)^3},$$
$$a^{k, \ell}_2 = \frac{4}{(R-1)^2}\left(\frac{ikr^2\;\text{Re} - 2\ell^2 - 1}{r^2}\right),$$
$$a^{k, \ell}_1 = \frac{2}{R-1} \left(\frac{ikr^2\;\text{Re} + 2\ell^2 + 1}{r^3}\right),$$
$$a^{k, \ell}_0 =  -\frac{\ell^2\left(ikr^2\;\text{Re} - \ell^2 + 4\right)}{r^4}.$$
Due to the homogeneous boundary conditions of Equations \eqref{Eq:Fk Bd 1} and \eqref{Eq:Fk Bd 2}, a polynomial interpolant is not needed. 
We limit the boundary conditions to two by defining 
\begin{equation}
f_{k, \ell}(s) = (1 - s^2)q_{k, \ell}(s).
\label{Eq:Second Order Numerical Equation 2}
\end{equation}
Then, we write Equation \eqref{Eq:Second Order Numerical Equation 1} as
\begin{eqnarray}
&& b^{k, \ell}_4 q_{k, \ell}^{(4)}(s) + b^{k, \ell}_3 q_{k, \ell}^{(3)}(s) + b^{k, \ell}_2 q_{k, \ell}^{(2)}(s)\nonumber\\
&&+ b^{k, \ell}_1 q_{k, \ell}^{(1)}(s) + b^{k, \ell}_0 q_{k, \ell}^{(0)}(s) = \text{RHS},
\label{Eq:Second Order Numerical Equation 3}
\end{eqnarray}
where 
$$b^{k, \ell}_4 = a^{k, \ell}_4(-s^2 + 1),$$
$$b^{k, \ell}_3 = -8s\; a^{k, \ell}_4 + a^{k, \ell}_3 (-s^2 + 1),$$
$$b^{k, \ell}_2 = -12\;a^{k, \ell}_4 - 6s\;a^{k, \ell}_3 + a^{k, \ell}_2(-s^2 + 1),$$
$$b^{k, \ell}_1 = -6\;a^{k, \ell}_3 - 4s\;a^{k, \ell}_2 + a^{k, \ell}_1(-s^2 + 1),$$
$$b^{k, \ell}_0 =  -2\;a^{k, \ell}_2 - 2s\;a^{k, \ell}_1 + a^{k, \ell}_0(-s^2 + 1).$$
Equation \eqref{Eq:Second Order Numerical Equation 3} is computed by using MATLAB's backslash command, and we obtain our solution $f_{k, \ell}(s)$ by plugging in $q_{k, \ell}(s)$ in Equation \eqref{Eq:Second Order Numerical Equation 2}.

\section{Numerical Validation of the Net Force}\label{Sec:App4}
From the asymptotic analysis of Section \ref{Sec:Low Amplitude Analysis}, we compute the steady solutions up to third order in amplitude and obtain the net force. We validate the results from our analysis by comparing with the numerical solution of the incompressible, time-dependent Navier Stokes equations on an annular domain with a horizontal velocity of strength \eqref{Eq:NonDim Far Field} on the outer cylinder generated using the Dedalus software\cite{burns2020dedalus}, which is  an open-sourced library that uses a parallelized framework and to solve PDE's with spectral methods. The velocity is computed on a Fourier-Chebyshev mesh with 256 Fourier modes and 512 Chebyshev points.  We solve for the flow to time $T = 105$ to allow the initial transient to decay and compute the steady flow by taking the time-average of the last 5 periods.

\begin{figure}[t]
    \centering
    \includegraphics[width=1\linewidth]{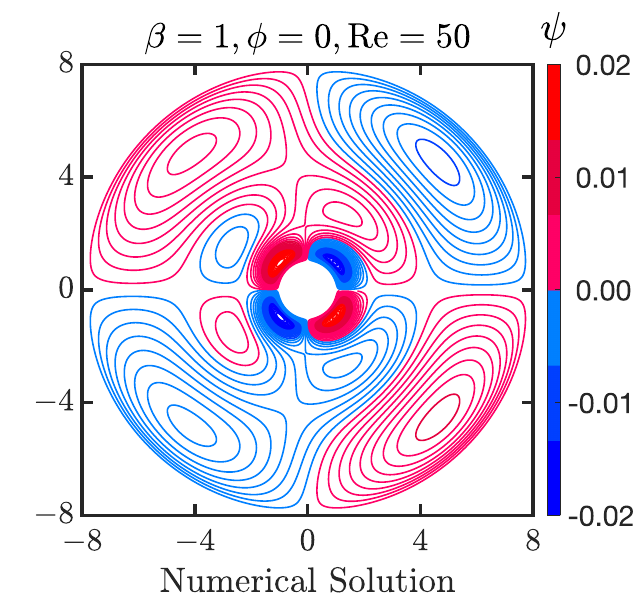}
    \caption{Contours of the steady stream function at amplitude ratio $\beta = 1$, phase difference $\phi = 0$, and Re = 10 from the numerical solution at amplitude ratio $\beta = 1$, phase difference $\phi = 0$, $\epsilon = \sqrt{0.02}\approx0.14$, and Re = 50} 
\label{Fig:Dedalus}
\end{figure}

Figure \ref{Fig:Dedalus} shows the stream functions computed from the numerical solution to the Navier Stokes equations at amplitude ratio $\beta = 1$, phase difference $\phi = 0$, $\epsilon = \sqrt{0.02}\approx0.14$, and Re = 50. The relative difference between the stream function from Figure \ref{Fig:Flow Structure}(b) is approximately 6.11\%, and the relative difference of the net force is about 12.45\%.   For smaller amplitude $\epsilon = \sqrt{0.0002}$, the relative error from the stream function and the net force is around 0.0618\% and 0.12\%, correspondingly.

\section{Effect of Domain Size}\label{Sec:App6}
We revisit the computational results in Section \ref{Sec:Computational Results} and \ref{Sec:Pumping and Net Force} for varying domain size $R$. The numerical computations were performed with 200 Chebyshev points to account for the larger domains.    

\begin{figure}
    \centering
    \includegraphics[width=1\linewidth]{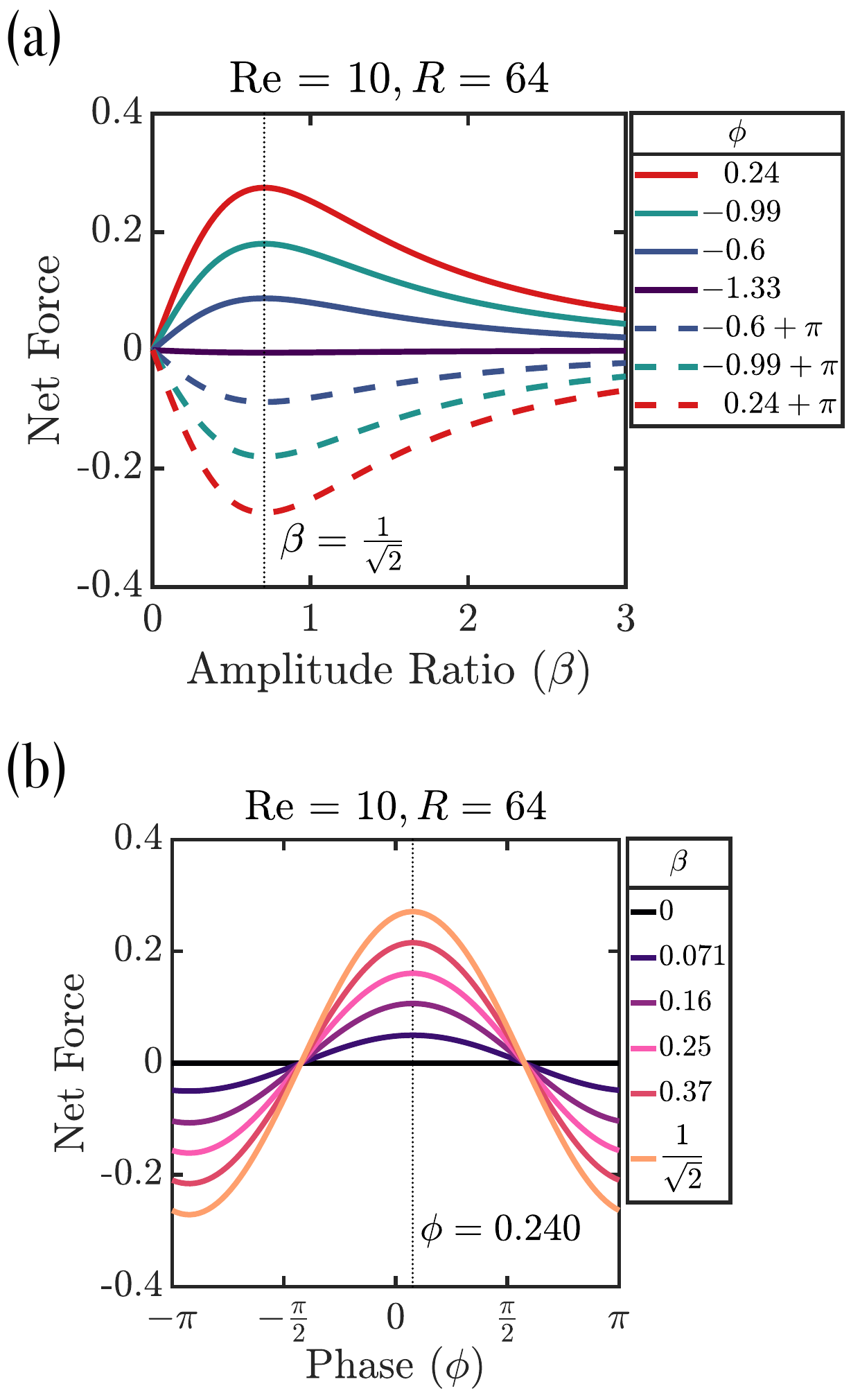}
    \caption{For domain size $R = 64$ and Re = 10, net force as a function of (a) amplitude ratio $0\leq \beta \leq 3$ for varying phase and (b) phase $-\pi\leq \phi \leq \pi$ for varying amplitude ratio.
    The vertical dashed lines show $\beta_{max} = 1/\sqrt{2}$ and $\phi_{max}=0.240$ for (a) and (b) respectively.} 
\label{Fig:Larger Domain 1}
\end{figure}

In Figure \ref{Fig:Larger Domain 1}, we show the net force for different amplitude ratios and phases differences for a domain of size $R=64$.  These plots closely resemble the corresponding results on domain size $R=8$ shown in Figure \ref{Fig:Net Force Beta and Phi}, with two notable differences: on the larger domain, the phase difference where the force is maximized is slightly reduced and the magnitude of the force is smaller.  

\begin{figure}
    \centering
    \includegraphics[width=1\linewidth]{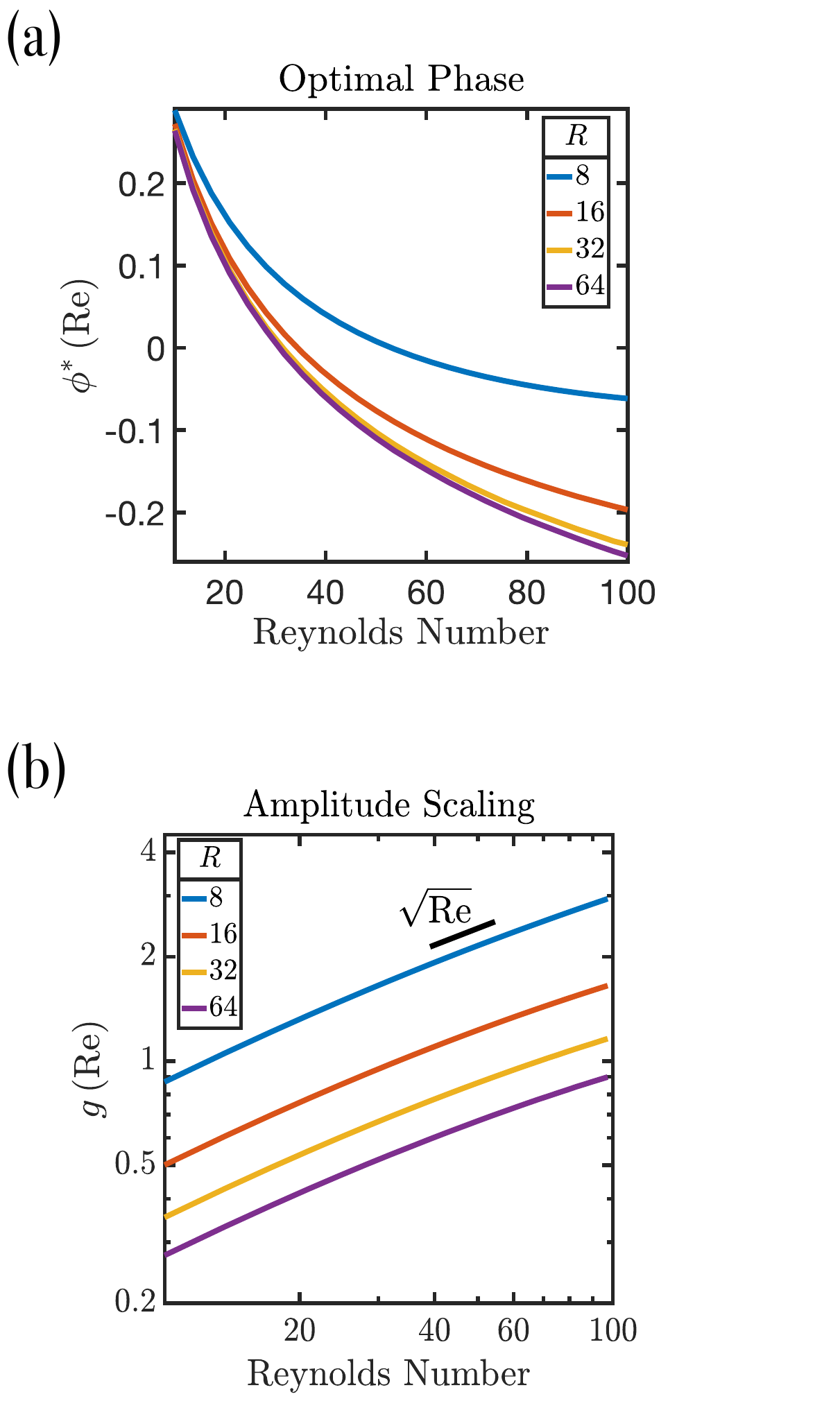}
    \caption{The optimal phase (a) $\phi^*(\text{Re})$ and maximum of the net force on the cylinder in the annulus (b) $g(\text{Re})$ as a function of the Reynolds number for domain sizes $R = 8, 16, 32,$ and $64$.} 
\label{Fig:Larger Domain 2}
\end{figure}

Figure \ref{Fig:Larger Domain 2}(a) displays the optimal phase difference as a function of Reynolds number, $\phi^*(\text{Re})$, for different domain sizes. The qualitative changes of  $\phi^{*}$ with Re are similar on different domains. The optimal phase difference decreases with domain size, with a more substantial change at higher Re.  The maximum force amplitude as a function of Reynolds number is shown in Figure \ref{Fig:Larger Domain 2}(b) for different domain sizes. It shows the same approximate $\sqrt{\text{Re}}$ scaling on different domains, but the size of the force decreases with increasing domain size.

\begin{figure}
    \centering
    \includegraphics[width=1\linewidth]{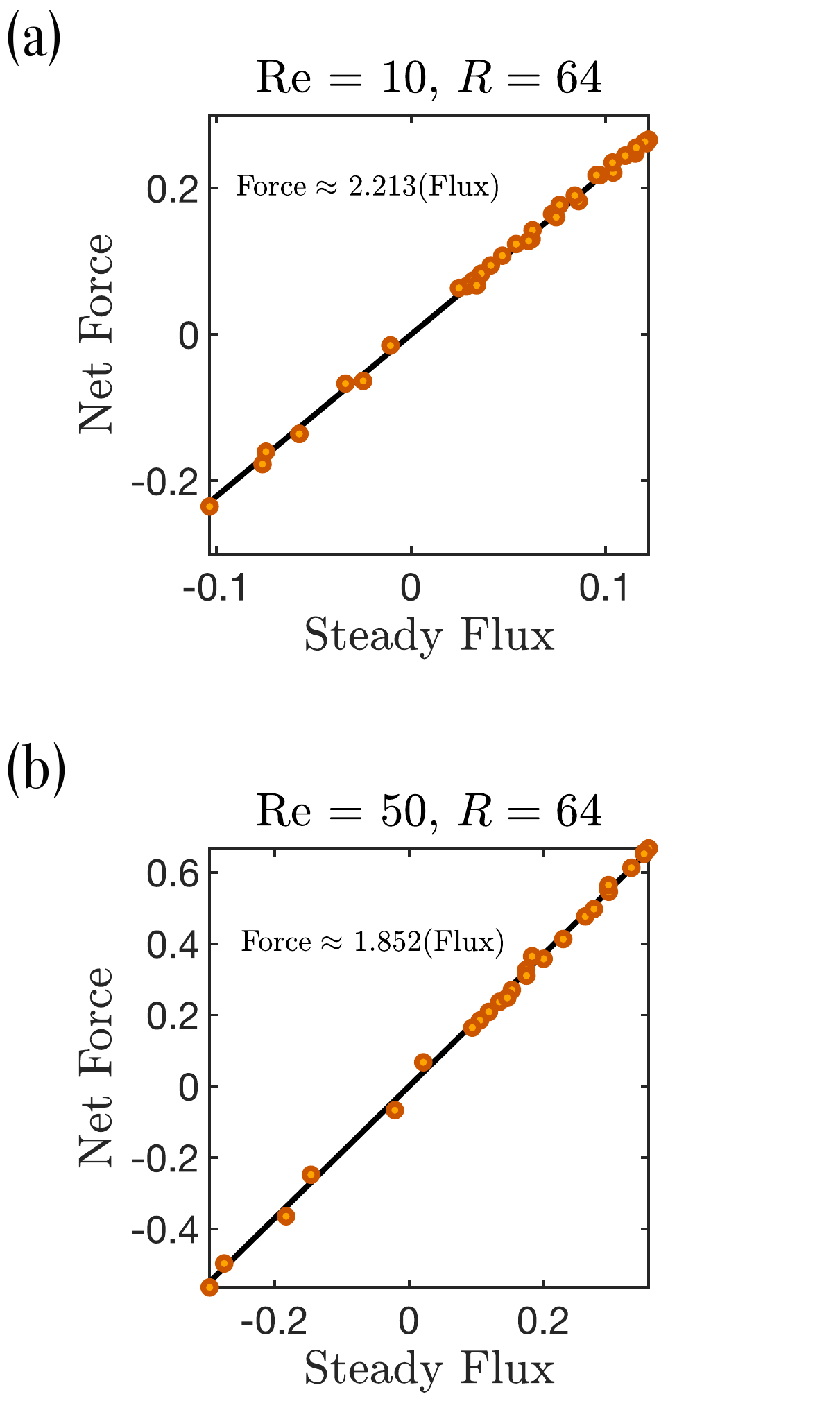}
    \caption{At $R = 64$, the net force on the cylinder computed on the annular domain is proportional to the steady time-averaged flux through a channel for $\epsilon = 0.5$ for (a) Re = 10 and (b) Re = 50. The solid line is the best fit line through the origin.} 
\label{Fig:Larger Domain 3}
\end{figure}

Finally, we show that the proportionality constant between the net force on the cylinder in the annular domain from our analysis and the time-averaged flux through a channel from our simulations depends on the domain size.
For the numerical simulation at $\epsilon = 0.5$ and Re = 10 and Re = 50, we compute the net force from the analysis for the same amplitude, phase, and Reynolds number. 
Figure \ref{Fig:Larger Domain 3}(a,b) shows the scatter plot of the net force against the steady flux, and comparing with Figure \ref{Fig:Beta and Phi Comparison}(c,f), we find that the slope of the best fit line through the origin depends on both the Reynolds number and the domain size.

\section{Numerical Method for Channel Simulation}\label{Sec:App5}
We use the Immersed Boundary (IB) method to solve the Navier-Stokes equations and simulate the flow around an oscillating cylinder that is inside a channel. 
The IB method uses an Eulerian coordinate system for the fluid and a Lagrangian coordinate system for the immersed structures (i.e.\ cylinder or channel walls) \citep{peskin2002immersed}.
 To begin, we define $s$ as the parametric coordinate of a structure and $\boldsymbol{X}(s, t)$ is its position.
 We use capital letters $\boldsymbol{X}(s, t)$, $\boldsymbol{U}(s, t)$, and $\boldsymbol{F}(s, t)$ to define position, velocity, and force density in Lagrangian coordinates, and similarly we use lower case $p(\boldsymbol{x}, t)$, $\boldsymbol{u}(\boldsymbol{x}, t)$, and $\boldsymbol{f}(\boldsymbol{x}, t)$ for pressure, velocity, and force density in Eulerian coordinates. 
 
We apply the forces on the structures to the surrounding fluid, and the fluid and structure move with the same velocity on the structure.
On the structure, the force density $\boldsymbol{F}(s, t)$ in Eulerian coordinates is given by
\begin{equation}
    \label{Eq:SpreadOperator}
    \boldsymbol{f}(\boldsymbol{x}, t) = S\boldsymbol{F} = \int_{\text{structure}}\boldsymbol{F}(s, t)\delta(\boldsymbol{x} - \boldsymbol{X}(s, t))\;ds,
\end{equation} 
where we define Dirac delta function as $\delta\left(\boldsymbol{x}\right)$. 
In \eqref{Eq:SpreadOperator}, the operator $S$ ``spreads'' the force density from the immersed structure to the surrounding fluid, and the fluid velocity is interpolated to the immersed structure by the adjoint of the spreading operator:
\begin{align}
    \label{Eq:InterpolatingOperator}
    \boldsymbol{U}(s, t) = S^*\boldsymbol{u} = \int_{\text{fluid}}\boldsymbol{u}(\boldsymbol{x}, t)\delta(\boldsymbol{x} - \boldsymbol{X}(s, t))\;d\boldsymbol{x}.
\end{align}
In our simulations, we prescribe the motion of the structure, and the force density on the structure, $\boldsymbol{F}(s, t)$, is determined implicitly by requiring the fluid velocity match the prescribed velocity of the boundary: 
\begin{align*}
    S^*\boldsymbol{u} = \boldsymbol{U}_b.
\end{align*}
We introduce the full system that describes the fluid and immersed boundaries as, 
\begin{align}
    \label{Eq:MomentumIB}
    \rho\left(\frac{\partial \boldsymbol{u}}{\partial t} + \boldsymbol{u}\cdot \nabla \boldsymbol{u}\right) 
        &= -\nabla p + \mu \Delta \boldsymbol{u} + S\boldsymbol{F},\\
    \label{Eq:IncompressibilityIB}
    \nabla \cdot \boldsymbol{u} &= 0,\\
    \label{Eq:Contraint}
    S^*\boldsymbol{u} &= \boldsymbol{U}_b,
\end{align}
where $\rho$ is the fluid density and $\mu$ is the fluid viscosity. 

We solve on a doubly periodic domain discretized into points equally spaced by $\Delta x$, and structures are discretized by points equally spaced by $\Delta s\approx\Delta x$. 
The grid resolution for channel simulations was $\Delta x=8/256$, so that there were $256$ grid points in the vertical direction, or equivalently $64$ points along the cylinder diameter. The time step was $\Delta{t}=5.00\times10^{-3}$.
For a detailed methodology and convergence of the numerical method, we refer to Appendix C of Lee \textit{et al.}\cite{lee2025pumping}

\bibliography{myCite}
\end{document}